\documentclass[aoas,preprint]{imsart}

\RequirePackage[OT1]{fontenc}
\RequirePackage{amsthm,amsmath}
\RequirePackage{natbib}
\RequirePackage[colorlinks,citecolor=blue,urlcolor=blue]{hyperref}

\RequirePackage{graphicx,multirow}

\startlocaldefs
\numberwithin{equation}{section}
\theoremstyle{plain}

\endlocaldefs

\begin{document}

\begin{frontmatter}
\title{Bayesian Latent Pattern Mixture Models for Handling Attrition
   in Panel Studies With Refreshment Samples}

\runtitle{Bayesian Latent Pattern Mixture Models}

\begin{aug}
\author{\fnms{Yajuan} \snm{Si}\thanksref{m1}
\ead[label=e1]{ysi@biostat.wisc.edu}},
\author{\fnms{Jerome P.} \snm{Reiter}\thanksref{m2}
\ead[label=e2]{jerry@stat.duke.edu}}
\and
\author{\fnms{D. Sunshine} \snm{Hillygus}\thanksref{m2}
\ead[label=e3]{hillygus@duke.edu}}
\runauthor{Si et al.}

\affiliation{University of Wisconsin-Madison\thanksmark{m1}
and Duke University \thanksmark{m2}}

\address{Department of Biostatistics \& Medical Informatics\\
Department of Population Health Sciences\\
University of Wisconsin-Madison\\
Madison, WI 53726\\
\printead{e1}}

\address{Department of Statistical Science\\
Duke University\\
Durham, NC 27708\\
\printead{e2}}

\address{Department of Political Science\\
 Duke University\\
 Durham, NC 27708 \\
\printead{e3}}
\end{aug}

\begin{abstract}
Many panel studies collect refreshment samples---new, randomly sampled respondents who complete the questionnaire at the same time as a subsequent wave of the panel. With appropriate modeling, these samples can be leveraged to correct inferences for biases caused by non-ignorable attrition. We present such a model when the panel includes many categorical survey variables. The model relies on a Bayesian latent pattern mixture model, in which an indicator for attrition and the survey variables are modeled jointly via a latent class model. We allow the multinomial probabilities within classes to depend on the attrition indicator, which offers additional flexibility over standard applications of latent class models. We present results of simulation studies that illustrate the benefits of this flexibility. We apply the model to correct attrition bias in an analysis of data from the 2007-2008 Associated Press/Yahoo News election panel study. 

\end{abstract}


\end{frontmatter}

\section{Introduction}

Many longitudinal or panel surveys, in which the same individuals are interviewed repeatedly at different points in time, suffer from panel attrition. For example, in the American National Election Study, 47\% of respondents who completed the first wave in January 2008 failed to complete the follow-up wave in June 2010. Such attrition can result in biased inferences when the attrition generates non-ignorable missing data; that is, the reasons for attrition depend on values of unobserved variables \cite[e.g.,][]{schluchte1982,brown1990,diggle1994,ibrahim99,scharfstein1999,olsen2005,behr2005,bhattacharya2008inference,hogandaniels}.

Unfortunately, it is not possible to determine whether the attrition
is ignorable or non-ignorable, nor the extent to which attrition
impacts inferences, using the collected data alone. Consequently,
analysts have to rely on strong and generally unverifiable assumptions
about the attrition process. Many assume that attrition is a missing
at random (MAR) process; for example, MAR assumptions underlie the use
of post-stratification to adjust survey weights
\cite[e.g.,][]{holt:smith:79,gelman:carlin:00,henderson2010sour} and
off-the-shelf multiple imputation routines to create completed
datasets
\cite[e.g.,][]{pasek2009determinants,honaker2010missing}. Others allow
for specific not missing at random (NMAR) processes, characterizing
the attrition with a selection model \citep{hw1979, brehm1993phantom,
  kenward,scharfstein1999} or pattern mixture model
\citep{littlepatmix,little1994,daniels2000,
  roy2003,kenwardmolenthijs,lin2004,roy2008}.

Many panel surveys supplement the original panel with refreshment
samples. These are cross-sectional, random samples of new respondents
given the questionnaire at the same time as a subsequent
wave of the panel. For example, refreshment samples are included in
the National Educational Longitudinal Study of 1988, which followed a
nationally representative sample of 21,500 eighth graders in two year
intervals until 2000 and refreshed with cross-sectional samples in
1990 and 1992. Overlapping or rotating panels, in which a new study
cohort completes their first wave at the same time a previous cohort
completes a second or later wave, offer equivalent information.

Refreshment samples offer information that can be utilized to correct
inferences for non-ignorable panel attrition
\citep{hirano1998,bartels1999panel,hirano2001,sekhoninfo,bhattacharya2008inference,deng2012}. In
particular, analysts can use an additive non-ignorable (AN) model,
which comprises a model for the survey variables coupled with a
selection model for the attrition process \citep{hirano1998}. The
selection model must be additive in the variables observed and missing
due to attrition so that model parameters are identifiable. 

Specifying the models for the survey variables and the attrition
indicator can be challenging, even when the data include only a modest
number of variables. Consider, for example, a multinomial survey
outcome modeled as a function of ten categorical predictors. It is
difficult to determine which interaction terms to include in the
model, especially in the presence of missing data due to attrition
\citep{erosheva:etal:toulouse,milca2008,si:reiter:12}. The
model specification task is even more complicated when the analyst
seeks to model all survey variables jointly, for example, with
a log-linear model or sequence of conditional models (e.g., specify $f(a)$, then $f(b \mid a)$, then
$f(c \mid a, b)$, and so on). Joint modeling can be useful when the
survey variables suffer from item nonresponse. 

Recognizing this, \cite{si:reiter:hillygus:pa12} propose to use a Dirichlet process mixture of products of
multinomial distributions \citep{dunsonxing2009,si:reiter:12} to model the survey variables.
This offers the analyst the potential to capture complex dependencies among variables without selecting interaction effects, as well as to handle item nonresponse among the survey variables. However, for the attrition indicator model, \citet{si:reiter:hillygus:pa12} use probit regression with only main effects for the survey variables, eschewing the task of selecting interaction effects. While convenient, using a main-effects-only specification makes assumptions about the attrition mechanism that may not be realistic in practice. Furthermore, probit regressions can suffer from the effects of separability and near co-linearity among predictors \citep{gelman2008}, which complicates estimation of the AN model.

In this article, we present an alternative approach for leveraging refreshment samples based on Bayesian latent pattern mixture (BLPM) models. We focus on models for categorical variables. The key idea is to use the Dirichlet
process mixture of products of multinomial distributions for the
survey variables and attrition indicator jointly, thus avoiding
specification of an explicit selection model. We note that several
other authors \cite[e.g.,][]{brown03,roy2003,lin2004} have proposed
using mixture models for handling attrition outside of the context of
refreshment samples. As we show, the refreshment sample enables us to
allow the multinomial vectors within mixture components to depend on
attrition indicators, thereby encoding a flexible imputation engine
that reduces reliance on conditional independence assumptions. 

We were motivated by attrition in   
the Associated Press/Yahoo 2008 Election Panel  (APYN) study, a multi-wave
longitudinal survey designed to track the attitudes and opinions of
the American public during the 2008 presidential election campaign. 
The APYN study was the basis of dozens of news stories during the 
campaign and subsequent academic analyses of the election 
in the years since. However, 
the study lost more than one third of the original sample to
attrition by the final wave of data collection, calling into question
the accuracy of analyses based on the complete cases.  The APYN
included a refreshment sample in the final pre-election wave of data collection,
which we leverage via the BLPM model to create attrition-adjusted,
multiply imputed datasets.
We use the multiply imputed data to examine dynamics of public
opinion in the 2008 presidential campaign. 

The remainder of the article is organized as follows. In
Section~\ref{apyn-intro}, we introduce the APYN data.  In
Section~\ref{apm}, we describe pattern mixture models for refreshment
samples, including conditions under which model parameters are
data-identified. To our knowledge, this is the first description of
pattern mixture models in this context. In Section~\ref{blpm}, we
propose and motivate the BLPM model for refreshment sample
contexts. In Section~\ref{simulation}, we illustrate properties of the
BLPM model with simulation studies. Here, we demonstrate the benefits
of allowing the multinomial vectors within mixture components to
depend on attrition indicators. In Section~\ref{real}, we analyze the
American electorate in the 2008 presidential election, using the BLPM model to account for attrition in the APYN data. Finally, in Section~\ref{conclusion} we
summarize and discuss future research directions. 

\section{Description of APYN Data}
\label{apyn-intro}

\begin{table}[t]
\centering
\caption[Collected variables]{APYN variables from wave 1 (W1), wave 2
  (W2) and refreshment sample (Ref), with rates of item nonresponse. Item nonresponse arises either from refusals to answer the
question (respondent proceeded to the next question without giving a
response) or selection of a ``Don't know enough to say''
response. We note that 1,011 of the wave 1 participants attrited from
the panel by wave 2, which could result in attrition bias.}
  \label{varpm1}
  \begin{tabular}{lcrrr}
 \hline
 && \multicolumn{3}{c}{Item nonresponse counts ($\%$)}\\
 Variable  &Levels& W1: 2735&W2: 1724&Ref: 464 \\
 \hline
 Obama favorability&2&550 (20.1)&95 (5.5)&20 (4.3)\\
 Party identification (Dem., Rep., Ind.)&3&13 (0.5)&&9 (1.9)\\
 Ideology (Lib., Mod., Con.)&3&57 (2.1)&&10 (2.2)\\
 Age (18--29, 30--44, 45--59, 60+)&4&0&&0\\
 Education ($\leq$ HS, Some coll., Coll.)&3&0&&0\\
 Race (White, Non-white)&2&0&&0\\
 Gender&2&0&&0\\
 Income (Ks) ($<$30, 30--50, 50--75, $\geq$75)&4&0&&0\\
 Married indicator &2&0&&0\\
  \hline
 \end{tabular}
\end{table}

The APYN study included eleven waves of data
   collection and three refreshment samples spanning the 2008 primary
   and general U.S. election season. The survey was sampled from the
   GfK Knowledge Panel, which is one of the nation's only online,
   probability-based respondent pools designed to be statistically
   representative of the U.S. population.  The respondent pool is
   recruited via a probability-based sampling method using published
   sampling frames that cover 96\% of the U.S. population.  Sampled
   non-internet households are provided with a laptop and free internet
   service.  Individuals in the respondent pool are then invited to
   participate in online surveys, such as the APYN panel survey. 
   Surveys from the GfK KnowledgePanel are
   approved by the Office of Management and Budget for government
   research and have been used in hundreds of academic publications
   spanning diverse disciplines, including health and medicine,
   psychology, social sciences, public policy, and survey and
   statistical methodology.  More
   information about the survey methodology can be found at
   \url{http://www.knowledgenetworks.com/ganp/election2008/index.html}.

Wave 1 of the APYN was fielded on November 2, 2007 and was completed by 2,735 respondents
out of 3,548 contacted individuals. After the initial wave, these wave
1 respondents were invited to participate in each follow-up wave, even
if they failed to respond to the previous one. Consequently,
wave-to-wave attrition rates or completion rates vary across the
study. Three external refresh cross-sections were also collected: a
sample of 697 new respondents in January, 576 new respondents in
September, and 464 new respondents in October. Each of
  the refreshment samples is a random and cross-sectional sample of
 the GfK respondent pool. Our analysis focuses on 
wave 1 (November 2007) and the ninth wave with a corresponding
refreshment sample (October 2008, the final wave before the election),
which we label wave 2 for presentational clarity. As shown in Table~\ref{varpm1}, of those who
completed wave 1, 1,011 (37\%) respondents failed to complete the
October wave.  In previous research using the APYN data
\citep{pasek2009determinants,henderson2011dynamics,iyengar2012affect,henderson2010sour}, 
scholars have mostly relied on post-stratification weights to correct
for potential panel attrition bias, although 
\cite{pasek2009determinants} used standard multiple imputation via
\texttt{Amelia II} \citep{amelia}. \cite{deng2012} outline the
limitations of such approaches---both assume that the attrition is
MAR. 

The primary outcome of interest in pre-election polls tends to be
evaluations of the candidates, as analysts attempt to gauge levels of
candidate support within the electorate. Which candidate is most
likely to win the election? Who in the electorate supports each side?
Because the earliest waves of the APYN took place before the ballot
match-up was known---i.e., before Obama and McCain had been selected
as their party nominees---we focus on Obama favorability (coded as
favorable or not). This variable offers exact
comparability in question wording across survey waves and is highly
correlated with eventual vote choice (the tetrachoric correlation of
the items in wave 2 is 0.97). In examining Obama favorability, we
consider standard covariates from the voting behavior literature.
These include demographic variables  (from ``Age'' to ``Marital
status'' in Table~\ref{varpm1}) previously shown to be related to
candidate evaluations and/or panel attrition
\citep{frankel:hillygus:13pa}.\footnote{Demographic and political
  profile variables are collected in profile surveys when a panelist
  joins the KnowledgePanel and are updated continually; thus, they
  have few missing values for any one study.} We also consider two
relevant political background variables (``Party
identification'' and ``Ideology'' in Table~\ref{varpm1}) that are
typically considered time invariant in the context of a single
election cycle \citep{bartels11}.

\section{Additive Pattern Mixture Models for Refreshment Samples}
\label{apm}

Before introducing the BLPM model and analyzing the APYN data, we review 
the AN model of \cite{hirano1998} and present
a corresponding pattern mixture model formulation.
Suppose the data comprise a two wave panel of $N_p$ individuals with a
refreshment sample of $N_r$ new individuals in the second wave. For
all $N= N_p + N_r$ individuals, the data include $q_0$ time-invariant
variables $X=(X_1, \dots, X_{q_0})$, such as demographic or frame
variables. Let $Y_{1}=(Y_{11},\dots, Y_{1q_1})$ be the $q_1$ survey
variables of interest collected in wave 1. Let $Y_2=(Y_{21},\dots,
Y_{2q_2})$ be the corresponding $q_2$ survey variables collected in
wave 2. Here, we assume that $Y_1$ and $Y_2$ comprise the same
variables collected at different waves, although this is not
necessary. Among the $N_p$ individuals, $N_{cp} < N_p$ provide at
least some data in the second wave, and the remaining
$N_{ip}=N_p-N_{cp}$ individuals drop out of the panel. The refreshment
sample includes only $(X, Y_2)$; by design, $Y_1$ are missing for all
the individuals in the refreshment sample. In this section, we presume that
$X$, $Y_1$ in the panel, and $Y_2$ in the refreshment sample are not
subject to nonresponse, although we relax this when analyzing the
APYN.

For each individual $i=1,\dots, N$, let $W_i=1$ if individual $i$
would remain in wave 2 if included in wave 1, and let $W_i=0$ if
individual $i$ would drop out of wave 2 if included in wave 1.  
Here, $W_i$ is an indicator of panel attrition 
conditional on participation in wave 1; it is not an indicator of
item or unit nonresponse among individuals in the refreshment sample.
We note that $W_i$ is fully observed for all individuals in the panel but is
missing for the individuals in the refreshment sample, since
individuals in the refreshment sample are not provided the chance to
respond in wave 1. Putting it all together, the concatenated data have
the structure illustrated in Table~\ref{graph}.

 \begin{table}[t]
\centering
\caption[Data Structure of Panel and Refreshment Samples.]{Structure
  of panel and refreshment samples. Notation for sample sizes in
  parentheses. The total number of individuals in both datasets is $N=N_p + N_r$.}
\label{graph}
\begin{tabular}{r|c|c|c|}
\cline{2-4}
&Time-Invariant&Wave 1& Wave 2 \\
\cline{2-4}
\multirow{2}{*}{Panel ($N_p$)}&\multirow{3}{*}{$X$}
&\multirow{2}{*}{$Y_{1}$} & $Y_{2}$, $W=1 \,\,\,\,\,\,\,\,\,(N_{cp})$\\
\cline{4-4}
&&&\textcolor{red}{$Y_{2}$=?}, $W=0 \,\,\,(N_{ip})$\\
\cline{3-4}
Refreshment Sample ($N_r$) & & \textcolor{red}{$Y_1$=? }& $Y_{2}$, $\textcolor{red}{W=?} $\\
\cline{2-4}
\end{tabular}
\end{table}

The AN model requires a joint model for $(Y_1, Y_2 \mid X)$ and a selection model for $(W \mid X, Y_1, Y_2)$, that is,
\begin{eqnarray}
	\nonumber \label{Y} (Y_1, Y_2) \mid X   &\sim& f(Y_1,Y_2|X, \Theta)\\
	\label{R} W \mid Y_1, Y_2, X  &\sim& f(W|X, Y_1, Y_2, \Theta),
\end{eqnarray}
where $\Theta$ generically represents the parameters for both models. To enable identification, (\ref{R}) must exclude interactions between $Y_1$ and $Y_2$.

As an example of an AN model, suppose $Y_1$ and $Y_2$ are binary variables and $X$ is empty, as in \cite{hirano1998}. One specification of the additive non-ignorable selection model is
\begin{align}
	\label{eq:AN1} Y_{i1} &\sim \mbox{Bern}(\pi_1), \quad \mbox{logit}(\pi_{1}) = \alpha_0 \\
	\label{eq:AN2} Y_{i2} \mid Y_{i1} &\sim \mbox{Bern}(\pi_{i2}), \quad \mbox{logit}(\pi_{i2}) =
\beta_0 + \beta_1Y_{i1} \\
	\label{eq:AN3} W_{i} \mid Y_{i1}, Y_{i2}  &\sim \mbox{Bern}(\pi_{iW}), \quad
\mbox{logit}(\pi_{iW}) =  \tau_0  + \tau_1 Y_{i1} + \tau_2 Y_{i2}.
\end{align}

For a pattern mixture model representation, we require a model for $(W
\mid X)$ and for $(Y_1, Y_2 \mid X, W)$, that is,
\begin{eqnarray}
	\nonumber \label{Wpm} W \mid X  &\sim& f(W|X, \Theta)\\	
	\nonumber(Y_1, Y_2) \mid X, W   &\sim& f(Y_1, Y_2|X, W, \Theta).
\end{eqnarray}
Using the basic example, one specification of the additive pattern mixture (APM) model is
\begin{align}
	 \nonumber W_{i}  &\sim \mbox{Bern}(\pi_{W}), \quad \mbox{logit}(\pi_{W}) =  \omega_0 \\
	 \nonumber Y_{i1} \mid W_i &\sim \mbox{Bern}(\pi_{i1}), \quad \mbox{logit}(\pi_{i1}) = \delta_0 +
\delta_1 W_i \\
	\label{Ypm} Y_{i2} \mid Y_{i1}, W_i &\sim \mbox{Bern}(\pi_{i2}), \quad \mbox{logit}(\pi_{i2}) =
\gamma_0 + \gamma_1 W_i + \gamma_2 Y_{i1},
\end{align}
which contains as many free parameters as in (\ref{eq:AN1}) --
(\ref{eq:AN3}) and thus is data-identified. To enable identification, we exclude interactions between $Y_1$ and $W$ in (\ref{Ypm}). We note that both the AN
and APM models can include interactions with $X$ and readily extend to
other data types.


\section{Bayesian Latent Pattern Mixture Models}
\label{blpm}

We now develop an APM model for categorical data with $q 
=q_0+q_1+q_2$ variables.
Let $Z=(X,Y_1,Y_2)=(Z_1,\dots, Z_q)$ comprise all
potentially collected variables. We order variables so that
$j=1, \dots, q_0$ for $X$ variables, $j = q_0+1,
\dots, q_0+q_1$ for $Y_1$ variables, and $j=q_0+q_1+1, \dots,
q$ for $Y_2$ variables. For $i=1,\dots,N$ and
$j=1,\dots,q$,  without loss of generality let $Z_{ij} \in
\{1,\dots, d_{j}\}$ denote the level of variable $j$ for unit $i$,
where $d_{j} \geq 2$ is the total number of levels for variable $j$.

We specify the pattern mixture model as $f(W)f(Z\mid W)$, including
$X$ in the joint distribution of the survey variables. This
facilitates imputation of (ignorable) item nonresponse in $X$, and
allows us to take advantage of computationally efficient latent class
representations of categorical data. Specifically, we adapt the
truncated Dirichlet process mixture of products of multinomial
distributions (DPMPM) developed by \cite{dunsonxing2009}, used
previously for multiple imputation of missing cross-sectional data by
\cite{si:reiter:12}. The DPMPM assumes that each
   individual is a member of a latent class, and that within each class
   the variables follow  independent multinomial
   distributions. Averaging the multinomial probabilities over the
   latent classes induces  global dependence among the variables.

For $i=1, \dots, N$, let $s_i \in \{1, \dots, K\}$ indicate the 
class of individual $i$, and let $\pi_h =\textrm{Pr}(s_i 
= h)$ where $h = 1,\dots,K$. We assume that $\pi 
= (\pi_1, \dots, \pi_K)$ is the same for all individuals.
For $j=q_0+1, \dots, q_0+q_1$, let $\psi_{hjz} = 
\mbox{Pr}(Z_{ij} = z | s_{i}=h)$ be the probability of 
$Z_{ij}=z$ for any value $z$ given that individual $i$ is in 
class $h$. For $j=1, \dots, q_0$ and $j=q_0+q_1+1, 
\dots, q$, let $\psi_{hjz}^{(1)} = \mbox{Pr}(Z_{ij} =z | 
W_i=1, s_{i}=h)$ and
$\psi_{hjz}^{(0)} = \mbox{Pr}(Z_{ij} = z | 
W_i=0, s_{i}=h)$ be the
probabilities of $Z_{ij}=z$ for any value $z$ given that 
individual $i$ is in class $h$ for each value of $W_i$. The complete-data 
likelihood for $(s_i, W_i, Z_{i})$ in the BLPM is as follows.
\begin{align}
  s_{i} \mid \pi &\sim \textrm{Multinomial}(\pi_1, \dots, \pi_K) 
\label{BLPM:s}\\
         W_i \mid s_i &\sim  \textrm{Bernoulli}(\rho_{s_{i}}) 
\label{BLPM:W}.
\end{align}
When $j \in \{q_0\textrm{+}1, \dots, q_0\textrm{+}q_1\}$, we have
\begin{equation}
         Z_{ij} \mid s_i \sim 
\textrm{Multinomial}(\{1,\dots,d_{j}\},\psi_{s_{i}j1},\dots,\psi_{s_{i}jd_{j}})\label{BLPM:XY1}.
\end{equation}
When $j\in \{1, \dots, q_0, q_0+q_1+1, \dots, q\}$, we have
\begin{align}
         Z_{ij}\mid s_i, W_i=1&\sim \textrm{Multinomial}(\{1,\dots,d_{j}\},\psi_{s_{i}j1}^{(1)},\dots,\psi_{s_{i}jd_{j}}^{(1)}) 
\label{BLPM:XY2W1}\\
         Z_{ij}\mid s_i, W_i=0&\sim \textrm{Multinomial}(\{1,\dots,d_{j}\},\psi_{s_{i}j1}^{(0)},\dots,\psi_{s_{i}jd_{j}}^{(0)}) 
\label{BLPM:XY2W0}.
\end{align}

The BLPM model is a mixture of pattern mixture models, where 
\begin{equation}
\nonumber \label{eq:jng-ci}
         f(Z_i, W_i)=\sum_{h=1}^{K}\mathrm{Pr}(s_i=h)f(W_i|s_i=h)f(Z_i|W_i,
s_i=h).
\end{equation}
As in the DPMPM, we assume that $(Z_{q_0+1}, \dots, Z_{q_0+q_1}$), that is, 
$Y_1$, follow independent, class-specific multinomial distributions that 
are also independent of $W$ (and $X, Y_2$). However, we depart from the 
DPMPM by letting
$(Z_{1}, \dots, Z_{q_0}, Z_{q_0+q_1+1}, \dots, Z_{q})$ follow
class-specific, independent multinomial distributions that depend on
$W$. Relaxing the conditional independence between $Y_2$ and $W$
(that is,  $Y_2$ is independent of $W$ within any latent class) is 
possible because of information offered
by the refreshment sample. We force $Y_1$ and $W$ to be independent
within latent classes to enable identification, following the
strategy outlined in Section~\ref{apm}. We allow $X$ to depend on $W$
within classes to offer additional flexibility for settings where the 
distributions of $X$ are substantially different across attriters and 
non-attriters. When this is not the case---the distributions of $X$ are 
observed for both $W=1$ and $W=0$---one  can specify the 
model so that $X$ does not depend on $W$ within classes, thereby reducing 
the number of parameters to estimate.

For the prior distribution on $\pi$, we use the stick-breaking representation of a Dirichlet process prior distribution \citep{seth94}, truncating at large $K$ for computational convenience. In particular, we have
\begin{align}
	\pi_h &= V_h\prod_{g<h}(1-V_g)  \label{BLPM:pi}\\
	 V_h &\sim \textrm{Beta}(1,\alpha), \textrm{   for } h=1,\dots,K-1, \textrm{ and } V_K=1 \label{BLMP:v}\\
	\alpha &\sim \textrm{Gamma}(a_{\alpha}, b_{\alpha}) \label{BLPM:alpha}.
\end{align}
We use uniform prior distributions on all $\psi$ and $\rho$
parameters.  We follow \citet{dunsonxing2009} and \citet{si:reiter:12} and
set $a_{\alpha}=b_{\alpha}= 0.25$.  Setting $a_{\alpha}+b_{\alpha}= 0.5$
represents a small prior sample size and hence vague specification, thereby allowing
the data to dominate the cluster allocations. In our simulations and the APYN analyses, results are not sensitive to reasonable
default choices of $(a_\alpha, b_\alpha)$. We
estimate the model using a blocked Gibbs sampler \citep{ishwaran2001};
see the online supplement for an outline of the algorithm. 
 
We set $K$ to be large enough to help the DPMPM to describe the joint
distribution reasonably well yet still offer fast computation. Using
an initial proposal for $K$, say $K=20$, analysts can examine the
posterior distributions of the number of classes with at least one
assigned observation across Markov chain Monte Carlo (MCMC) iterates
to diagnose if $K$ is large enough. When there is significant posterior mass at a
number of classes equal to $K$, the analyst should add more classes.
The analyst can repeat this diagnostic procedure until finding a
suitable $K$. We note that the posterior predictive distributions used
to generate imputations typically are very similar for any sufficiently
large $K$.

The usual truncated DPMPM model is based on (\ref{BLPM:s})--(\ref{BLPM:alpha}) but requires that $\psi_{hjc_j}^{(0)} =\psi_{hjc_j}^{(1)}$ in (\ref{BLPM:XY2W1}) and (\ref{BLPM:XY2W0}) for all $(h, j, c_j)$. This implies that all $Z$ are independent of $W$ within classes, which may not be the case. The refreshment sample offers information that allows us to relax this assumption, particularly for $Y_2$. Intuitively speaking, the refreshment sample offers information about $f(Y_2 \mid s)$, and the complete cases in the panel offer information about $f(Y_2 \mid s, W=1)$. These two distributions identify $f(Y_2 \mid s, W=0)$. Without the refreshment sample, we do not have information to differentiate $f(Y_2 \mid s, W=0)$ and $f(Y_2 \mid s, W=1)$; as a consequence, we are forced to make the unverifiable assumption of conditional independence between $Y_2$ and $W$. In Section~\ref{simulation}, we present simulation studies that illustrate the biases that can result in when falsely assuming the conditional independence assumption.

The model can be used for posterior inference or for multiple imputation. For the latter, analysts select $m$ of the 
sampled completed datasets after convergence of the Gibbs sampler. These datasets should be spaced sufficiently 
so as to be approximately independent. This involves thinning the MCMC samples so that the autocorrelations 
among parameters are close to zero. Multiple imputation inferences then can be based on all $N$ units in the 
concatenated data. Alternatively, as discussed in \citet{deng2012}, some statistical agencies or data analysts may prefer 
to disseminate or base inferences on only the original panel after using the refreshment sample for imputing the missing 
values due to attrition. This might be preferable when combining the original and freshened samples complicates interpretation 
of sampling weights and design-based inference. Additionally, using only the $N_p$ completed panel cases reduces 
sensitivity of inferences to the specification of the multiple imputation model, which enters the analysis only 
for completing $Y_2$ for the attriters.  As pointed out by reviewers of this article, survey-weighted analyses of the multiply imputed data  can result in biased estimates of variance \citep{kimkalton}.  This can result from lack of congeniality \citep{meng:1994} 
of the imputation model and survey-weighted analysis.

\section{Simulation Studies}
\label{simulation}

In this section, we present results of simulation studies that
illustrate the potential of the BLPM model to account for
non-ignorable attrition. We use two data generation mechanisms: one in
which $Y_2$ and $W$ are not independent within classes, and one in
which they are independent within classes. We compare the performance
of the BLPM model to the usual DPMPM, a model that assumes $Y_2$ and
$W$ are conditionally independent. In each scenario, we set
$N_p=2,000$ and $N_r=1,000$.  Each wave includes
$q_1=q_2=5$ binary variables; for simplicity, we do
not include any $X$ variables. Table~\ref{sim1-noci} displays the
values of $\pi$ and the $\psi$ parameters for each scenario. These
designs result in non-trivial dependence structures; for example, we
ran Pearson's chi-square tests in the true datasets and rejected
independence at the $0.05$ significance level for 29 out of the 45 paired combinations among the $10$ variables.

In each replication of the simulation, we generate a dataset with values of $(Z, W)$ for all $N=3,000$ records; we call this the true data. We delete the values of $Y_2$ for all records in the panel with $W_i=0$ and the values of $(Y_1, W)$ for all $N_r$ records in the refreshment sample. The resulting dataset has the structure in Table~\ref{graph} without $X$. We fit the BLPM and DPMPM models using the Gibbs sampler, imputing $Y_2$ in the panel when $W=0$ and $(Y_1, W)$ in the refreshment sample in each MCMC iteration. For each scenario, we run 100 independent replications of the simulation.

\begin{table}[t]
\centering
\caption[Component and marginal probabilities for data simulation]{Latent class and marginal probabilities for simulations. The first five $\psi$ parameters correspond to $Y_{1j}$ variables, and the last five $\psi$ parameters correspond to $Y_{2j}$ variables. The columns labeled ``marginal'' are the weighted averages of $\psi$ over the latent classes.}
\label{sim1-noci}
\resizebox{\columnwidth}{!}{%
\begin{tabular}{lcccccccc}
  \hline
& \multicolumn{4}{c}{$Y_2$ and $W$ not Cond. Ind.} & \multicolumn{4}{c}{$Y_2$ and $W$ are Cond. Ind.} \\Parameter & $h = 1$ & $h=2$ & $h=3$ &Marginal  & $h = 1$ & $h=2$ & $h=3$ &Marginal\\
  \hline
$\pi$ & 0.4 & 0.3 & 0.3 & --  &  0.4 & 0.3 & 0.3 & -- \\
$\rho_{h}$ & 0.80 & 0.95 & 0.60  &  0.78 & 0.80 & 0.95 & 0.60  &  0.78  \\
$\psi_{h,1,1}$ & 0.25 & 0.55 & 0.85 & 0.52 & 0.25 & 0.55 & 0.85 & 0.52 \\
$\psi_{h,2,1}$   & 0.20 & 0.50 & 0.80 & 0.47 & 0.20 & 0.50 & 0.80 & 0.47\\
$\psi_{h,3,1}$    & 0.15 & 0.45 & 0.75 & 0.42  & 0.15 & 0.45 & 0.75 & 0.42 \\
$\psi_{h,4,1}$    & 0.10 & 0.40 & 0.70 & 0.37 & 0.10 & 0.40 & 0.70 & 0.37 \\
$\psi_{h,5,1}$    & 0.05 & 0.35 & 0.65 & 0.32 & 0.05 & 0.35 & 0.65 & 0.32\\
$\psi_{h,6,1}^{(0)}$, $\psi_{h,6,1}^{(1)}$    & 0.76,  0.38 & 0.46, 0.58 & 0.16, 0.78 & 0.49, 0.56 & 0.38,  0.38 & 0.58, 0.58 & 0.78, 0.78 & 0.56\\
$\psi_{h,7,1}^{(0)}$, $\psi_{h,7,1}^{(1)}$    & 0.77, 0.41 & 0.47, 0.61 & 0.17, 0.81 & 0.50, 0.59 & 0.41, 0.41 & 0.61, 0.61 & 0.81, 0.81 & 0.59 \\
$\psi_{h,8,1}^{(0)}$, $\psi_{h,8,1}^{(1)}$   & 0.78, 0.44 & 0.48, 0.64 & 0.18, 0.84 & 0.51, 0.62  & 0.44, 0.44 & 0.64, 0.64 & 0.84, 0.84 & 0.62 \\
$\psi_{h,9,1}^{(0)}$, $\psi_{h,9,1}^{(1)}$   & 0.79, 0.47 & 0.49, 0.67 & 0.19, 0.87 & 0.52, 0.65 & 0.47, 0.47 & 0.67, 0.67 & 0.87, 0.87 & 0.65 \\
$\psi_{h,10,1}^{(0)}$, $\psi_{h,10,1}^{(1)}$    & 0.80, 0.50 & 0.50, 0.70 & 0.20, 0.90 & 0.53, 0.68 & 0.50, 0.50 & 0.70, 0.70 & 0.90, 0.90 & 0.68\\
   \hline
\end{tabular}%
}
\end{table}
\normalsize

To evaluate the potential of the BLPM and DPMPM models to correct for attrition, as well as to compare them with each other, we focus primarily on the completed data estimates of $\mbox{Pr}(Y_{2}=1)$ in the panel. Let superscript $r=1, \dots, 100$ index replications of the simulation, and let superscript $t=1, \dots, T$ index MCMC iterations, where $T$ is the number of MCMC iterations used in computation. For all $(r, t)$, and for $i=1, \dots, N$ and $j=1, \dots, q$, let $z_{ij}^{(rt)}$ be the value of $z_{ij}$ in replication $r$ and MCMC iteration $t$.  Here, if $j > q_1$, $z_{ij}^{(rt)}$ is an observed value for all panel cases with $W_i^{(r)}=1$ and is an imputed value for all cases with $W_i^{(r)}=0$. For any variable indexed by $j > q_1$, we compute
\begin{equation*}
	\bar{z}_j^{(rt)} = \sum_{i=1}^{N_p} I(z_{ij}^{(rt)} =
        1)/N_p,\,\,\,\,\,\,
	\tilde{z}_j^{(r)} = \textrm{Median }(\bar{z}_j^{(r1)}, \dots, \bar{z}_j^{(rT)}).
\end{equation*}
Let $\bar{z}_j^{(r,true)}$ be the value of $\mbox{Pr}(Z_j=1)$ for the panel in the true data associated with replication $r$. We then compute
\begin{eqnarray*}
	DIF_j &=& |\sum_{r=1}^{100} \tilde{z}_j^{(r)}/100 - \bar{z}_j^{(r,true)}|\\
	RMSE_j &=& \left(\sum_{r=1}^{100} (\tilde{z}_j^{(r)} - \bar{z}_j^{(r,true)})^2 / 100\right)^{0.5}.
\end{eqnarray*}
The larger $DIF_j$ and $RMSE_j$, the more inaccurate are the completed-data estimate in the panel.  We use only the panel and not the concatenated data to magnify the impact of the models on imputation of the missing data due to non-ignorable attrition. We also report values of $DIF_j$ and $RMSE_j$ for the BLPM and DPMPM models for the means of $W$ and $Y_1$ in the refreshment sample.  These are both fully imputed in the models.

For each simulation run, we run MCMC chains for both models with $K=10$ classes---we obtained very similar results with $K=20$ and $K=30$. We run the chains for 20,000 and 30,000 iterations for the BLPM and DPMPM models, respectively, which exploratory runs suggest as
sufficient for the chains to converge. We keep every tenth draw from the final 10,000 draws of each chain, leaving $T=1,000$ MCMC draws for inference. To initialize the chains, for all $h$ we set $\rho_h =
N_{cp}/N_p$; set all $\phi$ parameters equal to 0.5;  set $\alpha  = 1$; and, generate all $K\textrm{-}1$ initial values of $V_h$ from (\ref{BLMP:v}) using $\alpha = 1$.

\begin{figure}[t]
\centering
\begin{tabular}{cc}
    \includegraphics[height=2in,width=0.5\textwidth]{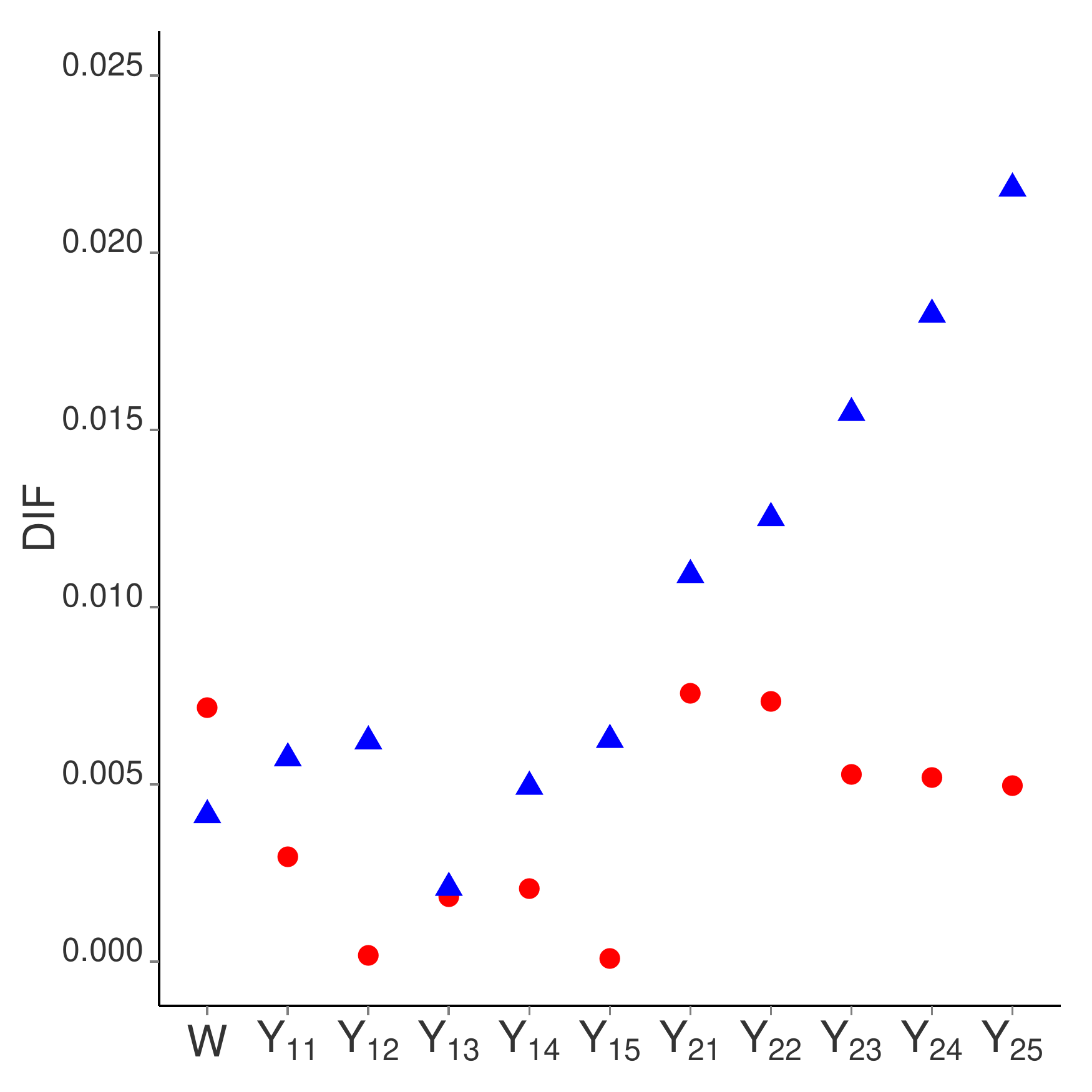}&
   \hspace{-0.2in}
    \includegraphics[height=2in,width=0.5\textwidth]{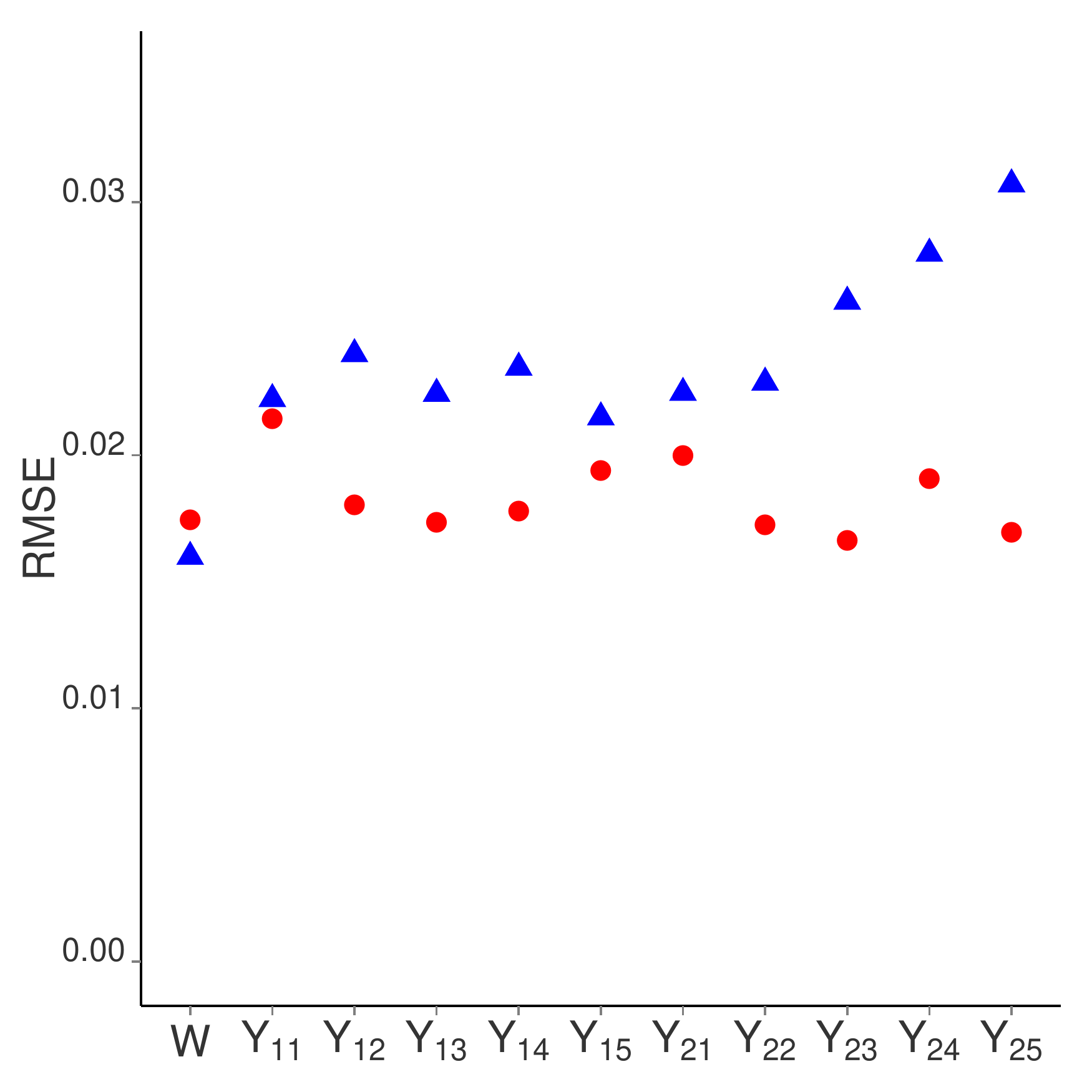}\\
\end{tabular}
\caption[Simulation outputs on datasets without conditional
independence between $Y_2$ and $W$.]{Simulation results
  when the data are generated with $Y_2$ and $W$ dependent within
  class. Results for DPMPM displayed with triangles and for BLPM with
  circles.}
\label{sm-rp}
\end{figure}

Figure~\ref{sm-rp} summarizes the values of $DIF_j$ and $RMSE_j$ for each quantity for both the BLPM and DPMPM models for the simulation with conditional dependence between $Y_2$ and $W$ within classes. We also computed the $DIF_j$ and $RMSE_j$ when estimating each $\mbox{Pr}(Y_{2j}=1)$ in the panel with only the complete panel cases. For this complete-cases estimator, the average values of $DIF$ and $RMSE$ across the 100 runs are shown in Table~\ref{sim1-noci-com}.

\begin{table}[t]
\centering
\caption[Results on the complete-cases estimator.]{Simulation results for the complete-cases estimator
  when the data are generated with $Y_2$ and $W$ dependent within
  class.}
\label{sim1-noci-com}
\begin{tabular}{lccccc}
  \hline
  $\mbox{Pr}(Y_{2j}=1)$&$j=1$&$j=2$&$j=3$&$j=4$&$j=5$ \\
$DIF_j$  &0.031&0.033& 0.039& 0.042& 0.046\\
 $RMSE_j$ &0.031& 0.033& 0.039&0.043& 0.047\\
  \hline
\end{tabular}
\end{table}
Compared to the results in Table ~\ref{sim1-noci-com}, the BLPM and DPMPM tend to offer smaller differences in point estimates, correcting the bias in complete-case analysis due to attrition. When estimating $\mbox{Pr}(Y_{2j}=1)$ using the panel data alone, the BLPM tends to be more accurate than the DPMPM.  The relative performance of the DPMPM worsens as the magnitude of the attrition bias increases, where by attrition bias we mean the difference in the marginal probabilities of $Y_{2j}$ for non-attriters and attriters, that is, $\sum_h \pi_h \psi_{hj1}^{(1)} - \sum_h \pi_h \psi_{hj1}^{(0)}$. We also tend to see better performance when predicting the missing $W$ and $Y_1$ in the refreshment sample, although the gaps are not as noticeable as those for $Y_2$.  For all $j>q_1$, the simulated matched pair standard errors are around $0.003$ for comparing $DIF_j$ for BLPM and DPMPM, and around $0.005$ for comparing $DIF_j$ for BLPM and the complete-case estimator.

\begin{figure}[t]
\centering
\begin{tabular}{cc}
    \includegraphics[height=2in,width=0.5\textwidth]{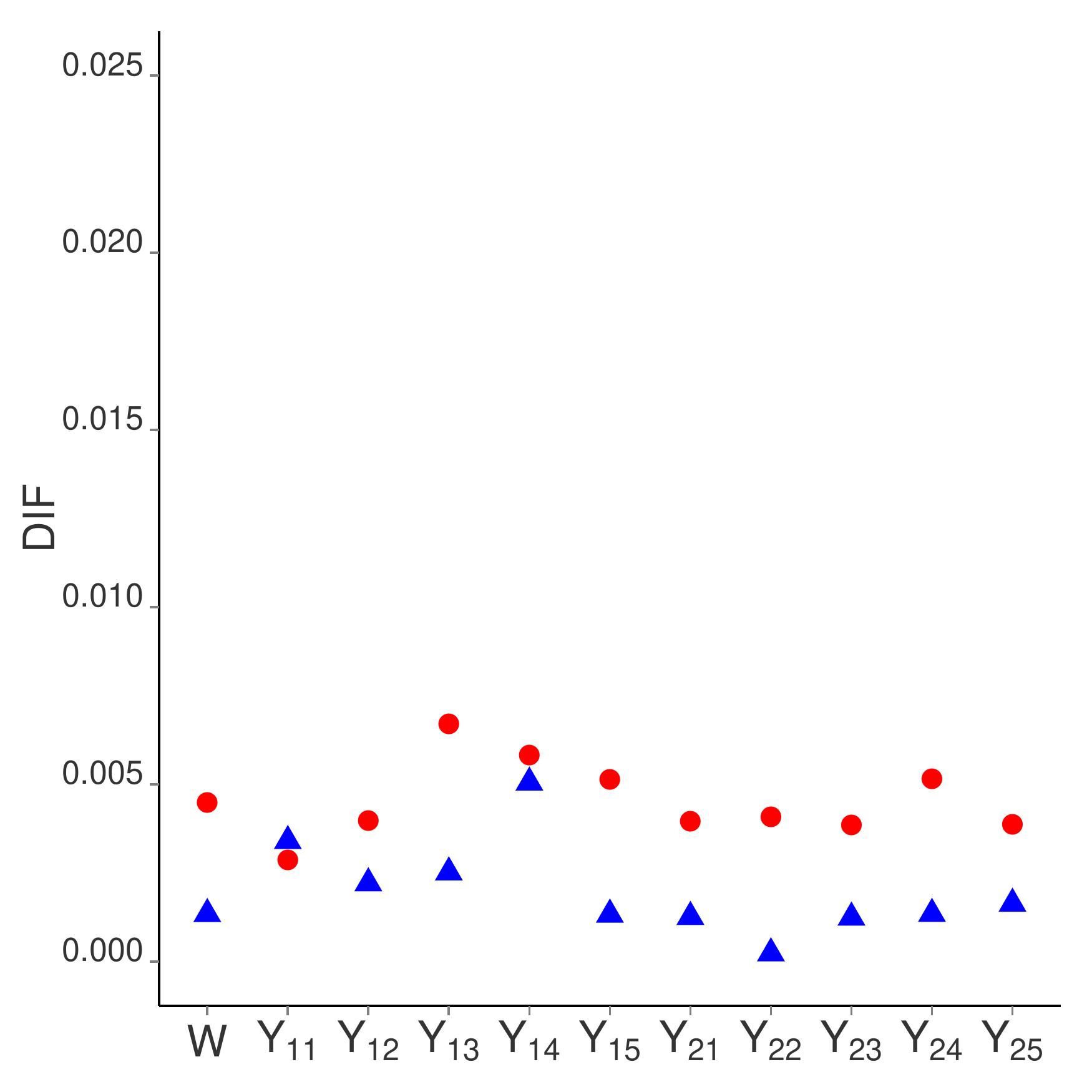}&
   \hspace{-0.2in}
    \includegraphics[height=2in,width=0.5\textwidth]{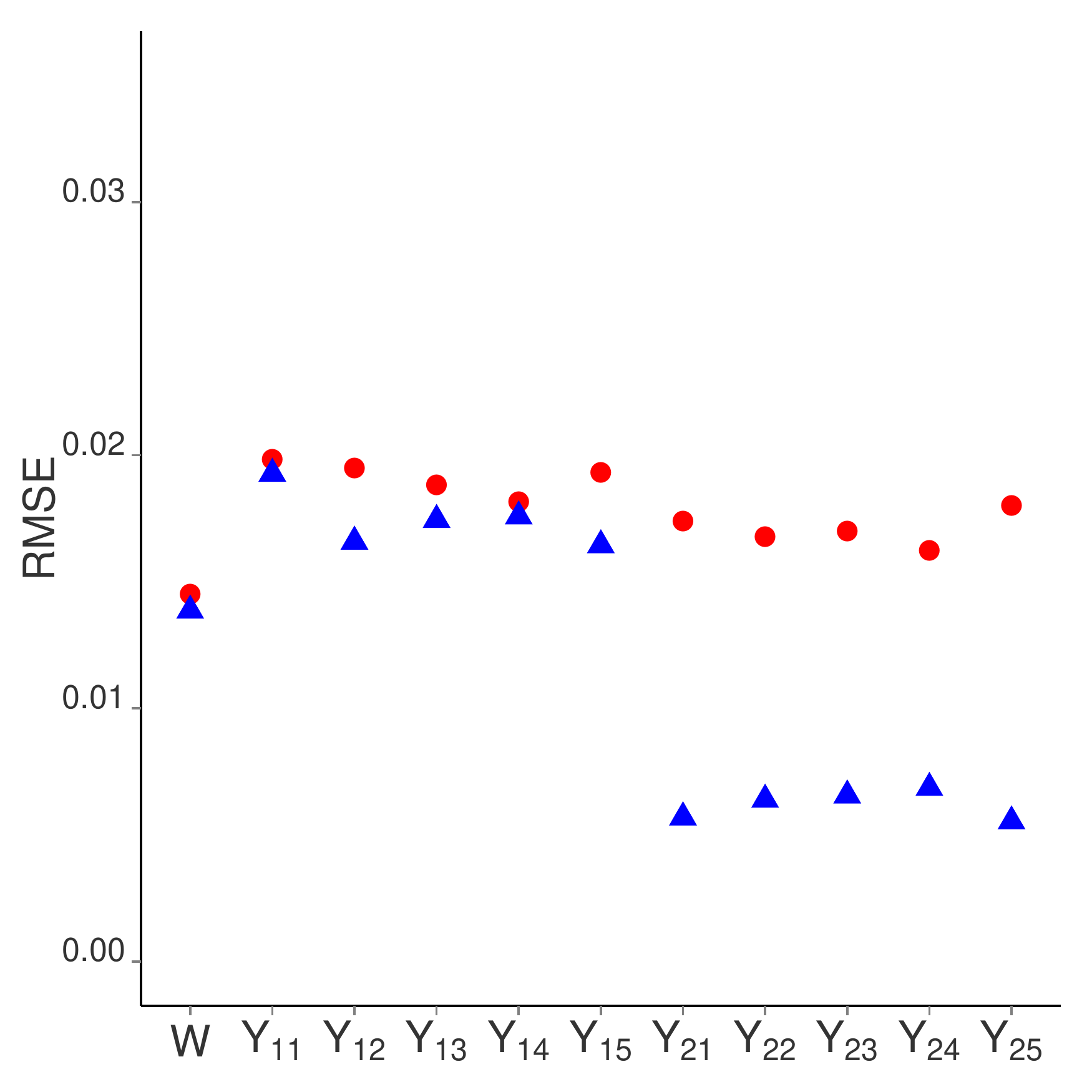}\\
\end{tabular}
\caption[Simulation outputs on datasets with conditional independence
between $Y_2$ and $W$.]{Simulation results when the data are generated with $Y_2$ and $W$ independent within class. Results for DPMPM displayed with triangles and for BLPM with
  circles.}
\label{sm-rp-ci}
\end{figure}

Figure~\ref{sm-rp-ci} summarizes the values of $DIF_j$ and $RMSE_j$ for each quantity for both the BLPM and DPMPM models for the simulation with conditional independence between $Y_2$ and $W$ within classes. For the complete-cases estimator, across the 100 runs, the average values of $(DIF_1, \dots, DIF_5)$ all equal approximately $0.016$ with associated $(RMSE_1, \dots, RMSE_5)$ equal to approximately $0.017$.  Once again, the BLPM and DPMPM tend to estimate each $\mbox{Pr}(Y_{2j}=1)$ using the panel data alone more accurately than the complete-case analysis. When estimating $\mbox{Pr}(Y_{2j}=1)$ using the panel data alone, the DPMPM tends to be slightly more accurate than the BLPM, but the differences are modest when compared to those in Figure~\ref{sm-rp}. The differences stem from estimating additional parameters in the BLPM, whereas the DPMPM has the exact specification. For all $j>q_1$, the simulated matched pair standard errors are around $0.002$ when comparing $DIF_j$ for BLPM and DPMPM, and $0.002$ when comparing $DIF_j$ for  BLPM and the complete-case estimator.

In summary, these simulation results suggest that both the BLPM and DPMPM can reduce attrition bias compared to 
using the complete cases. The BLPM is more flexible than the DPMPM in that it can protect against failure 
of the conditional independence assumption for $Y_2$ and $W$. However, when conditional independence holds, 
the BLPM estimates can be similar to those based on the DPMPM. A sensible default position with 
decent sample sizes is to use the BLPM, since the data do not inform whether conditional independence is appropriate.  

In our experience, in modest sample sizes both the BLPM and the DPMPM can suffer, as the latent class models
will sacrifice higher-order relationships among the variables.  Thus, it is crucial to check the fit of the models.  We suggest methods for doing so in the analysis of the APYN data (Section~\ref{real}).

\section{Using the BLPM to Correct for Attrition in the APYN data}
\label{real}

We now apply the BLPM model to account for attrition in the APYN data. To begin, we first provide 
some additional context on the survey design that is relevant for our imputations and analyses. 
Throughout, we refer to cross-sectional unit nonresponse as non-participation or refusal in the wave
 when an individual is initially surveyed; attrition happens when an
individual drops out after participating in a previous
wave. For example, the refreshment sample is subject to cross-sectional unit nonresponse
but not attrition, as these individuals are only surveyed at wave 2.

\subsection{Survey weights in the APYN}

The APYN data file includes survey weights at each wave. The wave 1 weights are the product of design-based 
weights and post-stratification adjustments for cross-sectional unit nonresponse at wave 1.    
These post-stratification adjustments assume the cross-sectional unit nonresponse is missing at random, as is common in the literature~\citep[e.g.,][]{hirano1998,bhattacharya2008inference,das2011nonparametric}.
The wave 2 weights for the 1724 panel participants include post-stratification adjustments for 
attrition in the panel, for cross-sectional unit nonresponse at wave 1, and for cross-sectional unit nonresponse among cases in the 
refreshment sample; the way that weights are reported does not allow us to disentangle these adjustments. 
Since we use the BLPM model to account for non-ignorable attrition, we disregard the wave 2 weights in all analyses.

The original panel is approximately an equal probability sample, with deviations due primarily to  
(i) slight oversampling of African American and Hispanic telephone exchanges 
and (ii) undersampling of areas where the MSN TV service network cannot be used and where there is no access
to the internet.  The post-strata in wave 1 are based on gender, race, the age groups in Table~\ref{varpm1}, the education groups in  
Table~\ref{varpm1}, census region, metropolitan area, and household internet access.
We include most of these variables in the BLPM model, thereby accounting for important aspects of the design when making imputations.  
The geographic variables and internet access are not strong predictors of Obama favorability given all the other
variables in Table~\ref{varpm1}. In a logistic regression with Obama favorability in wave 1 as the dependent variable, a drop in deviance test for the models with
and without  census region, metropolitan area, and internet access (including all other variables in $X$) results in
a p-value of 0.20.\footnote{We estimated the model with wave 1 data to
avoid any issues from non-ignorable attrition.}  Since these variables do not substantially improve our ability to predict 
the missing Obama favorability values, and are not of substantive interest in 
our analyses of the American electorate, we exclude them from the imputation model.

We use unweighted analyses to illustrate the attrition effects and describe the behavior of the BLPM model 
(as in Figures \ref{prob-x1} and \ref{prob-x2} in Section \ref{apynoutput}), and we use survey weighted analyses 
when computing finite population quantities (as in Figure \ref{prob-mar} in Section \ref{apynoutput}). 
The survey-weighted estimates account for the sampling design and cross-sectional unit nonresponse in wave 1 only. 
To make these estimates, we use the wave 1 weights for the 1724 panelists in multiple imputation 
inferences \citep{rubin:1987}.

\subsection{Generating Completed Datasets}
We run the BLPM with $K =30$ classes using the Gibbs sampler outlined in the online supplement, treating Obama favorability as $(Y_1, Y_2)$ and all other variables as $X$. As initial values for $W$ in the refreshment sample, we use independent draws from a Bernoulli distribution with probability $N_{cp}/N_p=0.63$. For missing data in $(X, Y_1, Y_2)$---due to item nonresponse and attrition---and $W$ in the refreshment sample, we implement the initialization steps of the MCMC as follows.
\begin{itemize}
	\item For any missing values in $X$, sample from the marginal distribution of $X$ computed from the observed cases in the combined panel and the refreshment sample.
	\item For any missing values in $Y_1$, sample from the observed marginal distribution of $Y_1$.	
	\item For missing values in $Y_2$ in the refreshment sample, sample from the observed marginal distribution of $Y_2$ in the refreshment sample.
	\item  For missing values in $Y_2$ in the panel for cases with $W_i=1$, sample from the observed marginal distribution of $Y_2$ in the panel.
	\item  For missing values in $Y_2$ in the panel for cases with $W_i=0$, sample from independent Bernoulli distributions with probabilities
$
	\mbox{Pr}(Y_2|W=0),
$	
obtained by $
[\mbox{Pr}(Y_2) - \mbox{Pr}(Y_2|W=1)\mbox{Pr}(W=1)]/\mbox{Pr}(W=0).
$
	Here, $\mbox{Pr}(Y_2)$ is estimated with the refreshment sample, $\mbox{Pr}(Y_2|W=1)$ is estimated with cases with $W_i=1$ in the panel, and $\mbox{Pr}(W=1)=0.63$.
\end{itemize}

For the initial values of the parameters, we set $\alpha=1$; set each $\rho_h = N_{cp}/N_p$; set each $\psi$ parameter equal to the corresponding marginal probability calculated from the initial
completed dataset; and set $V_h = 0.1$ for $h=1, \dots, K\textrm{-}1$. Each record's latent class indicator is initialized from a draw of a multinomial distribution with probability $\pi$ implied by the set of initial $\{V_h\}$.

We run the MCMC for 150,000 iterations, treating the first 100,000 as burn-in and thinning every 50th iteration. The trace plots of each variable's marginal probability suggest convergence. The posterior mode of the number of distinct occupied classes is 9, and the maximum is 18. This suggests that $K=30$ classes is sufficient. We collect $m=50$ completed datasets by keeping every twentieth
draw from the $T=1000$ thinned draws. We use only the $N_p$ records in the completed panels for multiple imputation inferences.

\subsection{Results}
\label{apynoutput}

\begin{table}[t]
\caption{\label{tab:frequencies}Unweighted percentages of respondents in each category in wave 1 and wave 2 of the panel (W1 and W2), and in the refreshment sample (Ref).  Percentages based on available cases only, before imputation of item nonresponse.}
\begin{center}
\begin{tabular}{lccc}
\hline
Variable & W1 & W2 & Ref.\\
\hline
Favorable to Obama & 0.553 & 0.549 & 0.617\\
Democrat & 0.327&0.318 &0.374 \\
Independent &0.369 &0.374 &0.312 \\
Liberal & 0.223&0.234  &0.289\\
Conservative & 0.366&0.370 &0.397 \\
Age 18--29 & 0.148&0.135 &0.110 \\
Age 30--44 &  0.284&0.284&0.213 \\
Age 45--59 &0.317 &0.320 &0.341 \\
HS Edu. or less & 0.343&0.325 &0.323 \\
College Edu. & 0.298&0.333 &0.308 \\
Non-white &0.230 &0.220 &0.177 \\
Female & 0.548& 0.537&0.565 \\
Income $<30K$ & 0.277&0.262 &0.170 \\
Income 30--50K & 0.269&0.270 &0.306 \\
Income 50--75K &0.225 & 0.235& 0.211\\
Married & 0.631&0.632 &0.647 \\
\hline
\end{tabular}
\end{center}
\end{table}

We begin by comparing the distributions of variables in wave 2 among the $N_{cp}$ non-attriters in the panel and the $N_r$ respondents in the refreshment sample; these are summarized in Table
\ref{tab:frequencies}. Among the non-attriters, 54.9\% favor Obama. In the refreshment sample, however,
61.7\% favor Obama. This suggests that people who liked Obama may have dropped out with higher frequency than those who did not.  As a sense of the magnitude of these differences, the 95\% confidence interval limits corresponding to these two percentages are (0.525, 0.573) and (0.572, 0.662), offering evidence that the difference may well be systematic. Of note, compared to the refreshment sample,
the $N_{cp}$ non-attriters are less likely to be Democrats and to be liberals,  more likely to be non-white and to have income below \$30,000, and more likely to be below age 45.

These differences in the marginal frequencies reflect the effects of attrition, as well as
differential cross-sectional unit nonresponse in the refreshment sample and initial wave. Reassuringly, national cross-sectional polls in October 2008 from Gallup, Fox News, and other major polling organizations also put Obama favorability ratings close to 62\% (\url{http://www.pollingreport.com/obama_fav.htm}),
suggesting the respondents in the refreshment sample faithfully represent Obama's favorability ratings at the time.
In our analyses, we assume that Obama favorability values missing for reasons other than attrition,
that is, due to cross-sectional item and unit nonresponse, are MAR given the variables in  the BLPM model.
Previous survey methodology research indicates that missingness mechanisms for attrition
  and cross-sectional nonresponse are distinct \citep[e.g.,][]{loosveldtetal97,groves:couper98, lynn05, groves06,
  smith:son10,olson2011}, so that one can plausibly consider attrition
as potentially non-ignorable even when assuming cross-sectional unit nonresponse is MAR.
See \citet{schifeling:hillygus:reiter} for further discussion of the effects on inferences of non-ignorable
cross-sectional unit nonresponse in the initial wave and refreshment sample.

\begin{figure}[h!]
\begin{tabular}{c}
	\includegraphics[width=\textwidth]{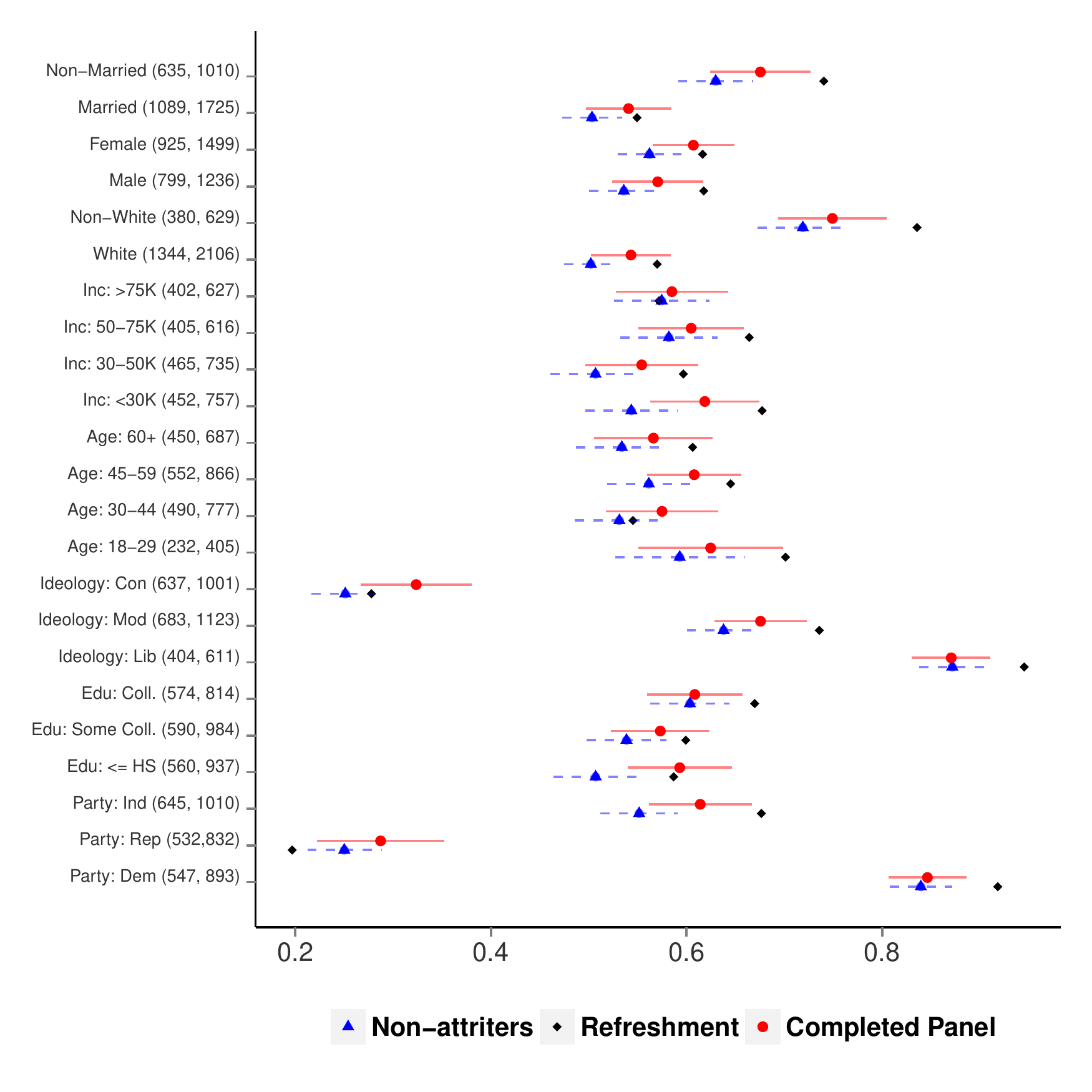}\\
\end{tabular}
 \caption[Conditional probability comparison on demographics between panel and refreshment samples]{Point estimates and 95\% confidence intervals for Obama favorability in various subgroups. Results presented for the $N_{cp}$ panel non-attriters, the $N_r$ refreshment samples, and the $N_p$ panel participants. Inferences based on unweighted analyses of the $m=50$ completed datasets, after multiple imputation of missing values via the BLPM model. The numbers in parentheses are the corresponding subgroup sizes, the first being the size among non-attriters and the second being among the completed panel. We randomly select one imputed dataset to obtain the sample sizes when the background variables are subject to item nonresponse.}
 \label{prob-x1}
\end{figure}

 Figure~\ref{prob-x1} displays estimated probabilities for Obama favorability for each of the subgroups defined by the time-invariant variables. For many subgroups, the estimates for non-attriters in the panel are noticeably different from those in the refreshment sample. This finding offers an important correction to the prevailing wisdom about the nature of panel attrition in political surveys. Research had previously concluded that attrition bias impacted outcomes related to political engagement (e.g., turnout) but not those related to candidate support (e.g., favorability) \citep{bartels1999panel,kruse2009panel}. The attrition biases within these subgroups provide evidence to the contrary. It is also noteworthy that the differences are most pronounced for women, low-income respondents, respondents aged 45--59, the least educated, and political independents. Many of these are the sub-populations often thought to lack a voice in American politics \citep{gilens2005inequality}, and these results suggest that panel attrition may further complicate accurate estimation of their political attitudes and preferences.

Figure~\ref{prob-x1} also reveals how the BLPM can correct for attrition bias. In particular, for most subgroups, the point estimate for the $N_p$ panel participants is shrunk towards the refreshment sample estimate; that is, the BLPM model corrects the bias due to attrition. The BLPM-corrected intervals tend to be wider than those computed with the non-attriters. This results from two sources of variability, namely the
estimation of the model parameters based on a modest-sized refreshment sample and the imputation of the $N_{ip}=1,011$ values of $Y_2$.

\begin{figure}[h!]
\centering
\includegraphics[width=\textwidth]{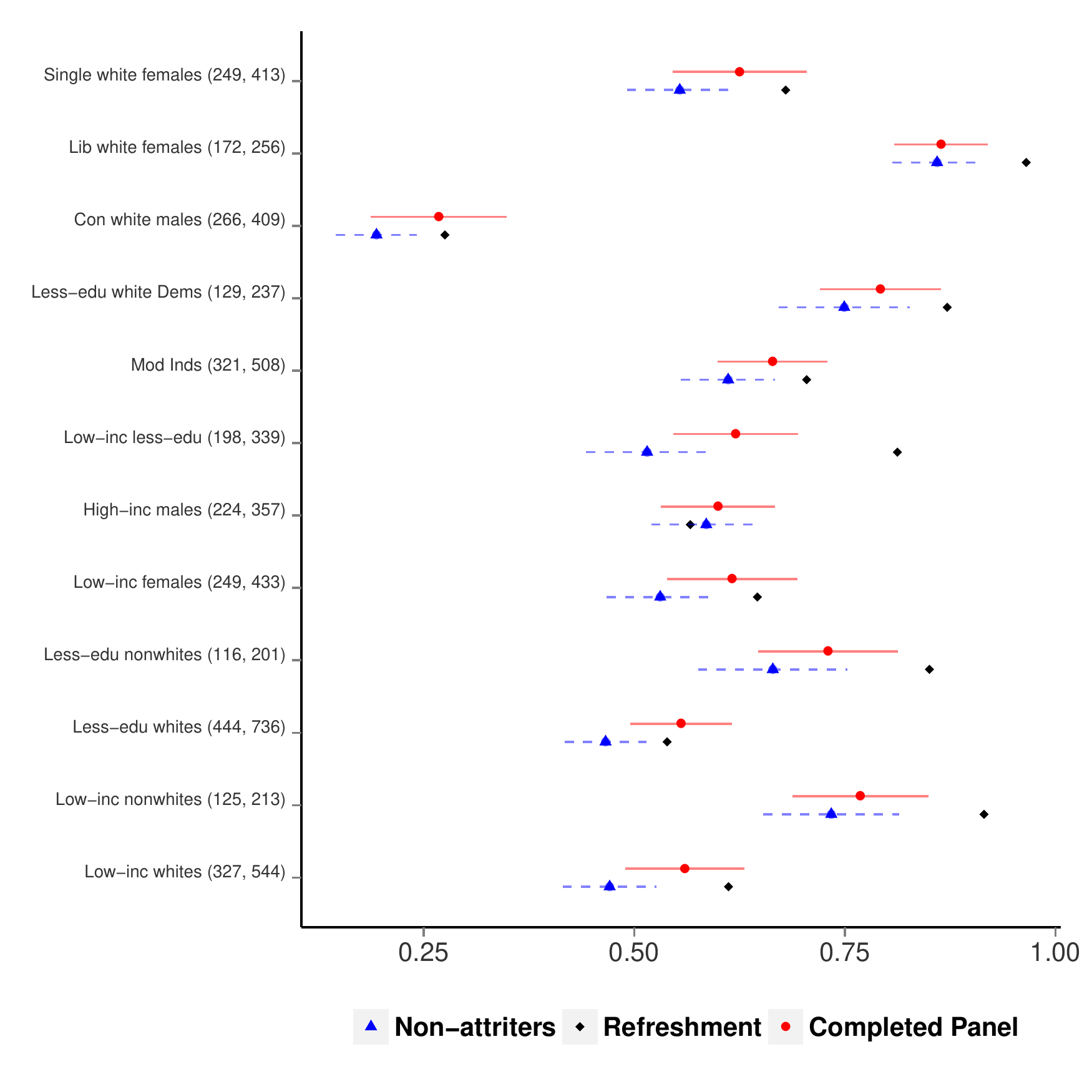}\\
 \caption[Conditional probability comparison on interacted demographics between panel and refreshment samples]{Point estimates and 95\% confidence intervals for Obama favorability in additional subgroups. Results presented for the $N_{cp}$ panel non-attriters, the $N_r$ refreshment samples, and the $N_p$ panel participants. Inferences based on unweighted analyses of the $m=50$ completed datasets, after multiple imputation of missing values via the BLPM model. The numbers in parentheses are the corresponding subgroup sizes, the first being the size among non-attriters and the second being among the completed panel. We randomly select one imputed dataset to obtain the sample sizes when the background variables are subject to item nonresponse.}
 \label{prob-x2}
\end{figure}

Figure~\ref{prob-x2} displays inferences for several smaller subgroups of substantive interest. Here, the BLPM's advantage over AN models is particularly prescient, as we are able to fit the BLPM model without having to specify (perhaps arbitrarily) a selection model with interaction effects. The attrition biases do appear to differ across the groups, suggesting the importance of using models that can capture interaction effects. Interestingly, high-income males appear not to experience substantial attrition bias, whereas various low-income and less educated groups appear to experience sizable underestimations of Obama favorability. As in Figure~\ref{prob-x1}, for most groups the BLPM generally shrinks point estimates towards those in the refreshment sample.

Of course, evaluating potential attrition bias is not the end goal of our analyses. Rather, having created attrition-adjusted imputations with the BLPM model, we now use the $m$ completed panel datasets to better understand the American electorate during the 2008 campaign. Here, we use survey-weighted analysis as follows. For each population percentage of interest and in each of the $m$ completed panel datasets, we compute the standard ratio estimate of the population percentage and the usual estimated variance based on the formula for unequal probability sampling with replacement \citep{lohr}. We obtain estimates with the \textit{survey} package \citep{surveyR} in $R$. We then combine the point and variance estimates using the multiple imputation rules \citep{rubin:1987}.

\begin{figure}
\centering
\begin{tabular}{c}
	\includegraphics[width=0.725\textwidth]{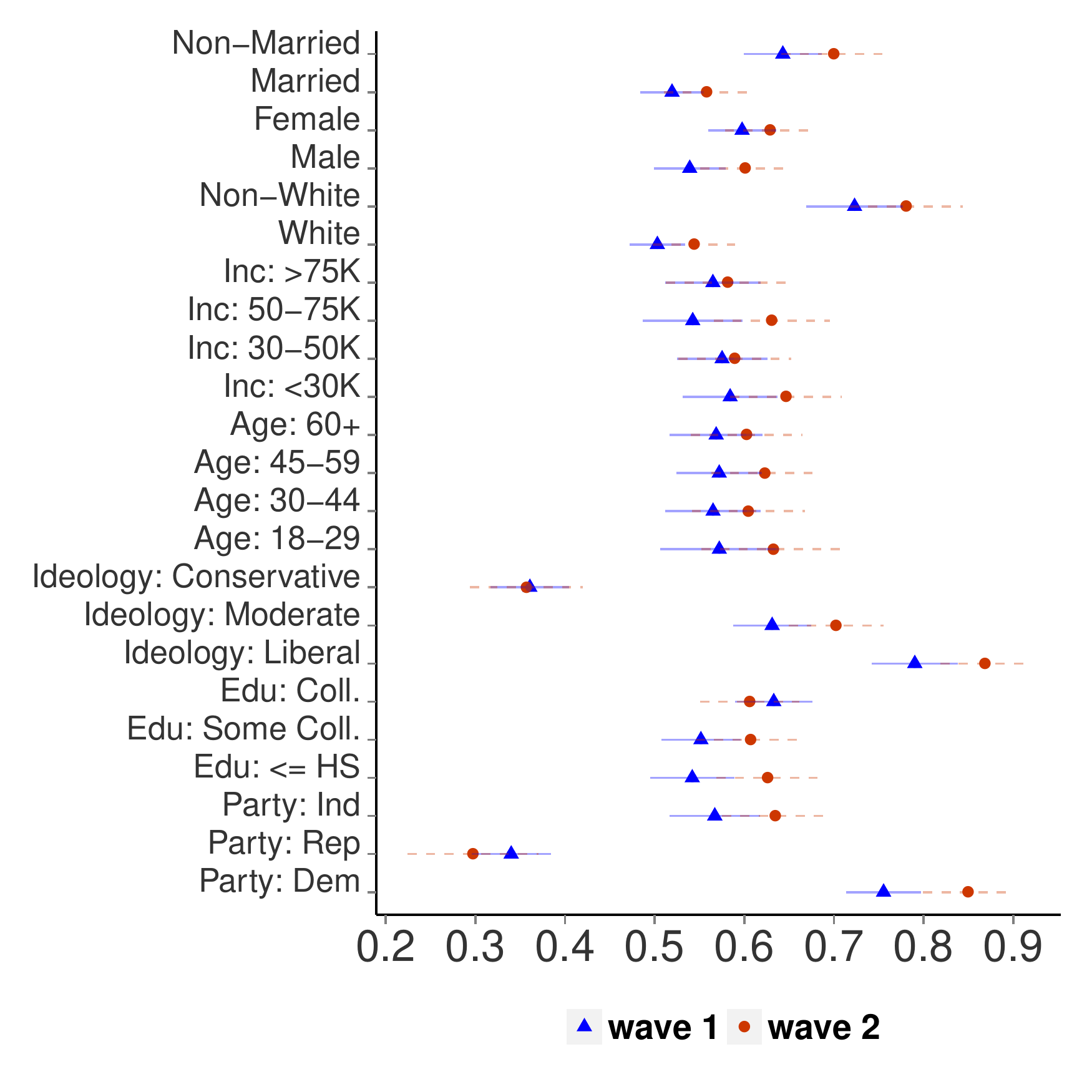}\\
	\includegraphics[width=0.725\textwidth]{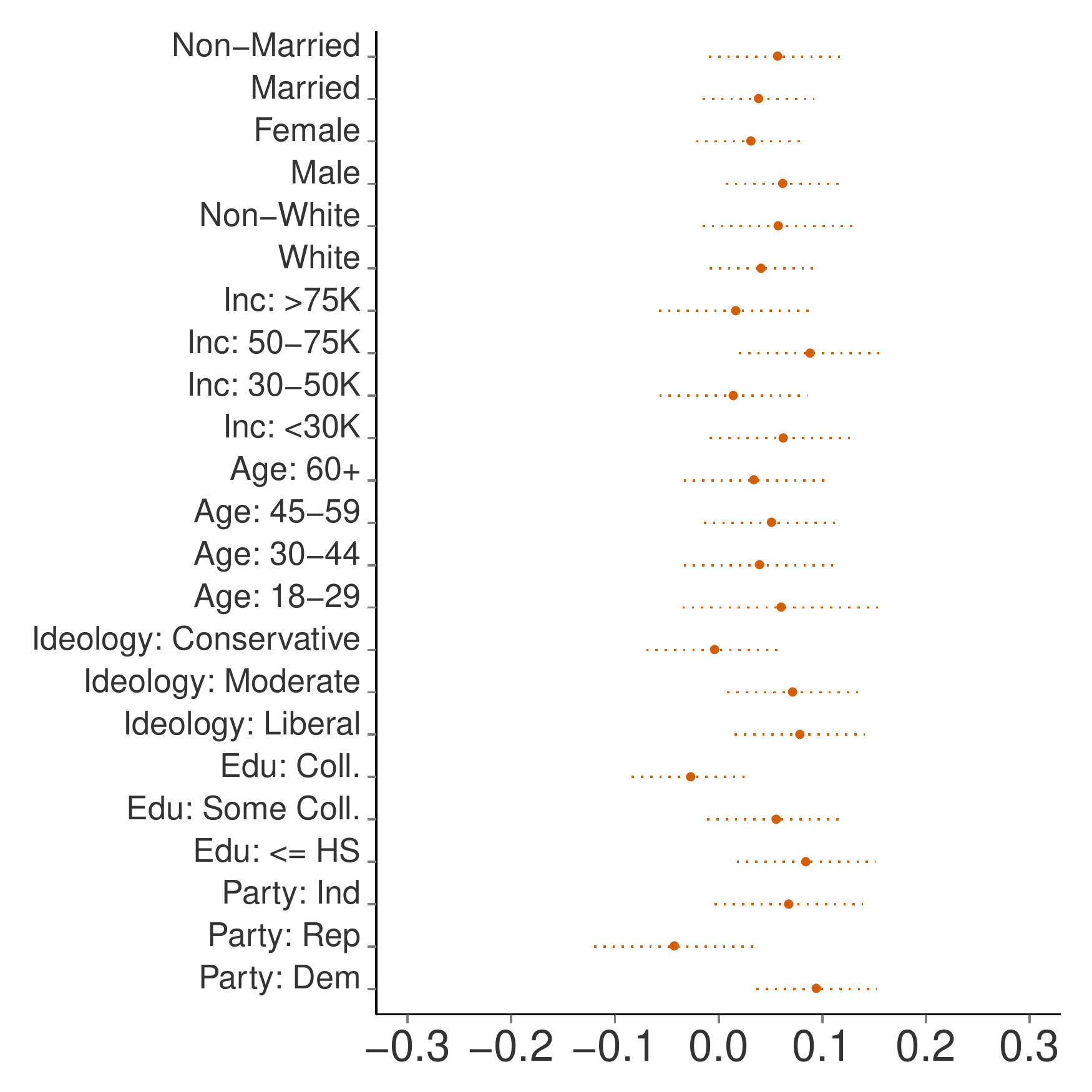}
\end{tabular}	
 \caption[Conditional probability comparison on demographics between wave 1 and wave 2]{Dynamics of Obama favorability ratings between wave 1 and wave 2. Top plot compares the marginal estimates in wave 1 and wave 2. Bottom plot presents the differences between wave 2 and wave 1. Results based on the $N_p$ panel participants after multiple imputation via the BLPM model. Inference based on survey-weighted estimation. }
 \label{prob-mar}
\end{figure}

Accounting for the wave 1 survey weights, the marginal estimate for Obama favorability in the last days before Election Day (wave 2) was $0.615$ $(0.576, 0.655)$, indicating Obama enjoyed the level of candidate support necessary to win the November election.  As can be seen in Figure~\ref{prob-mar}, Obama enjoyed higher levels of favorability among some expected subgroups---liberals, non-whites, and Democrats---in the weighted analysis for both waves. His high levels of favorability among other subgroups, especially moderates and Independents, offers the clearest signal of the likely election outcome.  It was only among self-reported Republicans and conservatives that Obama found favorability levels fall below 0.5. 

Comparing estimates across waves also suggests that the American electorate grew more favorable towards Obama as the campaign unfolded---the average marginal favorability in wave 1 is $0.569$ $(0.542, 0.597)$, as illustrated in Figure~\ref{prob-mar}. The increase in marginal favorability rating across waves is $0.046$ $(0.003, 0.089)$. In terms of attitude changes during the campaign among the various subgroups, most became slightly more favorable over time, 
with the exception of conservatives and Republicans who became slightly less favorable from wave 1 and wave 2. 
Most of these changes are not statistically significant due to sample size issues. The statistically significant changes in attitudes
are among Democrats, liberals, moderates, less educated, individuals with middle income, and males, who showed substantial increases in favorability towards Obama between wave 1 and wave 2. Overall, these patterns suggest that the partisan polarization in evaluations of Obama that characterize American
politics today actually started during the 2008 presidential campaign \citep{burden2009polls}.

We also fit the BLPM model assuming that $Y_1$ and $X$ are
conditionally independent of $W$ within latent classes. Reassuringly,
the conclusions from this version of the BLPM are similar to those
presented previously.

For comparison, we fit two additional models: the DPMPM model
described in Section~\ref{simulation} that does not have $Y_2$ depend
on $W$, and a MAR imputation model based on the DPMPM \cite[as
in][]{si:reiter:12} that disregards $W$ entirely. The results for both
models, reported in Section 3 of the online supplement, are similar to
each other but different from the BLPM results. These two alternative
models generally result in point estimates quite similar to those from
the non-attriters; in other words, they suggest that panel attrition
bias in Obama favorability is ignorable. This seems implausible given
the differences in Obama favorability seen in the non-attriters and
the refreshment samples. 

We also fit the semi-parametric AN model
  of \cite{si:reiter:hillygus:pa12}, which assumes a probit
  regression for $W$ conditional on $(X, Y_1, Y_2)$ and a DPMPM model for $(X,
  Y_1, Y_2)$.  Results are reported in Section 4 of the online
  supplement.  Both the semi-parametric AN and BLPM models suggest that the attrition is
  non-ignorable. Point estimates for the quantities in
  Figure~\ref{prob-x1} and~\ref{prob-x2} differ slightly; however, the differences are modest
  relative to the multiple imputation variances.  We prefer
  the BLPM results, as the model diagnostics of Section \ref{ppp-mi} suggest that the BLPM fits
  the data more effectively than the semi-parametric AN model. We
  further note that the semi-parametric AN model is computationally
  more intensive than the BLPM, as the probit regression for $W$ requires 
auxiliary data augmentation and Metropolis steps that are not necessary in the BLPM.

\subsection{Model Diagnostics}
\label{ppp-mi}

To  check the fit of the models, we follow the advice in \cite{deng2012} and use posterior predictive checks
\citep{meng1994,gelman:2005,he2010,burgreit10}. We use the BLPM model to generate $T^0=500$ data sets with no missing data in $(X, Y_1, Y_2, W)$, randomly sampling from the $T{=}1000$ available completed datasets. Let $\{D^{(1)}, \dots, D^{(T^0)}\}$ be the collection of the $T^0$ completed datasets. For each $D^{(t)}$, we also use the model to generate new values of $Y_2$ for all cases in the panel, including cases with $W_i=1$, and in the refreshment sample. This can be done after running the MCMC to convergence as follows. For given draws of parameter values and any item missing data in $(X,Y_1)$, sample new values for the observed and imputed $Y_2$ using the distributions in
the online supplement. Let $\{R^{(1)},\dots, R^{(T^0)}\}$ be the collection of the $T^0$ replicated datasets.

We then compare statistics of interest in $\{R^{(1)},\dots, R^{(T^0)}\}$ to those in $\{D^{(1)}, \dots, D^{(T^0)}\}$. Specifically, suppose that $S$ is some statistic of interest, such as a marginal or conditional probability in our context. For $t=1, \dots, T^0$, let $S_{R^{(t)}}$ and $S_{D^{(t)}}$ be the values of $S$ computed from $R^{(t)}$ and $D^{(t)}$, respectively. We compute the two-sided posterior predictive probability,
\begin{equation*}
	ppp = \frac{2}{T^0} * \textrm{min}\left(\sum_{t=1}^{T^0}I(S_{R^{(t)}}-S_{D^{(t)}} > 0), \sum_{t=1}^{T^0}I(S_{D^{(t)}}-S_{R^{(t)}} > 0)\right).
\end{equation*}
When the value $ppp$ is small, for example, less than $5\%$, this suggests the replicated datasets are systematically different from the observed dataset, with respect to that statistic. When the value of
$ppp$ is not small, the imputation model generates data that look like the completed data for that statistic. Recognizing the limitations of posterior predictive probabilities \citep{bayarri:berger}, we interpret the resulting $ppp$ values as diagnostic tools rather than as evidence from hypothesis tests that the model is ``correct.''

As statistics, we select $\mbox{Pr}(Y_{2}=1)$ in the refreshment sample, $\mbox{Pr}(Y_{2}=1\mid W=1)$ in the panel, $\mbox{Pr}(Y_{1}=1,
Y_{2}=1\mid W=1)$ in the panel, and $\mbox{Pr}(Y_{2}=1\mid X, W=1)$ in the panel for all conditional probabilities involved in the subgroup analyses in Figure~\ref{prob-x1} and~\ref{prob-x2}. This results in 38 quantities of interest. A histogram of the 38 values of $ppp$ is displayed in Section 4 in the online supplement. The analysis does not reveal any serious lack of model fit as none of the ppp values are below 0.20.

We repeat the same model diagnostics on the semi-parametric AN model
of~\cite{si:reiter:hillygus:pa12}.  Many of posterior predictive probabilities are
uncomfortably small. We believe the differences in the semi-parametric
AN and BLPM models result because the predictor function in the AN model for $W$ 
used by ~\cite{si:reiter:hillygus:pa12} includes only main effects, whereas the BLPM model does not {\em a priori} enforce a model for attrition.

\section{Concluding Remarks}
\label{conclusion}

The proposed Bayesian latent pattern mixture model offers a flexible way to leverage the information in refreshment samples in categorical 
datasets, helping to adjust for bias due to non-ignorable attrition. We have used this approach in analyzing the APYN study to better understand 
the preferences of the American electorate during the 2008 presidential campaign. Our findings suggest that panel attrition biased downward 
estimates of Obama favorability among many subgroups in the electorate. With a more accurate assessment of voter attitudes, we find that 
Obama had sufficiently high levels of favorability among key subgroups---independents and moderates---to suggest that the election outcomes
 were not really in doubt by late October.

The BLPM approach has key advantages over existing applications of additive non-ignorable models.  The BLPM avoids the difficult tasks of specifying a binary regression model for the attrition process. Unlike standard latent class models, the BLPM fully utilizes the information in the refreshment sample by allowing for conditional dependence within latent classes between wave 2 variables and the attrition indicator.
We note that a wide range of existing surveys have data structure amenable to BLPM modeling, including 
the General Social Survey, the 2008 American National Election Study, the Survey of Income and Program Participation, and the National Educational Longitudinal Study, to name just a few.

As with other modeling strategies for refreshment samples, the validity of the BLPM depends on several overarching assumptions. First, the initial wave of the panel and the refreshment sample should be representative of the same population of interest. Put another way, the units in the target population should
not change substantially between wave 1 and wave 2, although certainly the distributions of the substantive variables can do so. Second, any unit (or item) nonresponse other than that due to attrition is missing at random. Third, to ensure identifiability, we assume conditional independence between wave 1 survey variables and the attrition indicator within classes. 
When this assumption is unreasonable, the BLPM model---and any additive pattern mixture model---could fail to correct 
for attrition bias. Unfortunately, the data do not provide information about the plausibility of this assumption. 
Methods for assessing the sensitivity of results to violations of this assumption, as well as to violations of the two representativeness assumptions, are important areas for research.

\section*{Acknowledgements}
This research was supported by NSF grants SES-10-61241 and SES-11-31897.


\begin{supplement}
\sname{Supplement A}\label{suppA}
\stitle{Bayesian Latent Pattern Mixture Models for Handling Attrition in Panel Studies With Refreshment Samples}
\slink[doi]{COMPLETED LATER BY THE TYPESETTER}
\sdescription{The supplement includes the MCMC algorithms for the BLPM and DPMPM models, additional analyses of the APYN data using the DPMPM model and semi-parametric AN model, and details of the BLPM model diagnostics.}
\end{supplement}

\bibliographystyle{imsart-nameyear}
\bibliography{blpm}


\end{document}


\begin{frontmatter}
\title{Supplement A: Bayesian Latent Pattern Mixture Models for Handling Attrition
   in Panel Studies With Refreshment Samples}
\runtitle{Supp A: Bayesian Latent Pattern Mixture Models}

\begin{aug}
\author{\fnms{Yajuan} \snm{Si}\thanksref{m1}
\ead[label=e1]{ysi@biostat.wisc.edu}},
\author{\fnms{Jerome P.} \snm{Reiter}\thanksref{m2}
\ead[label=e2]{jerry@stat.duke.edu}}
\and
\author{\fnms{D. Sunshine} \snm{Hillygus}\thanksref{m2}
\ead[label=e3]{hillygus@duke.edu}}

\runauthor{Si et al.}

\affiliation{University of Wisconsin-Madison\thanksmark{m1}
and Duke University \thanksmark{m2}}

\address{Department of Biostatistics \& Medical Informatics\\
Department of Population Health Sciences\\
University of Wisconsin-Madison\\
Madison, WI 53726\\
\printead{e1}}

\address{Department of Statistical Science\\
Duke University\\
Durham, NC 27708\\
\printead{e2}}

\address{Department of Political Science\\
 Duke University\\ 
 Durham, NC 27708 \\
\printead{e3}}

\end{aug}
\end{frontmatter}

This online supplement comprises five sections. Section~\ref{ps-jblpm-ci} and Section~\ref{ps-jblpm} present
the Markov chain Monte Carlo algorithms for a Dirichlet process mixture of products of multinomial (DPMPM) model and for the Bayesian latent pattern mixture (BLPM) model. The former model imposes conditional independence between $(X,Y_2)$ and $W$ within
classes. Section~\ref{apyn-dpm} presents the analyses of the AP Yahoo News (APYN) election panel data when using the DPMPM model rather than BLPM model. Section~\ref{supp-ppp-an} presents results of an analysis of the APYN data using the semi-parametric additive non-ignorable model of \cite{si:reiter:hillygus:pa12}, including corresponding posterior predictive checks. Section~\ref{supp-ppp} includes the posterior predictive checks for the BLPM model. Throughout we use the same notation as in the main text.

\section{Posterior Computation Algorithm for DPMPM}
\label{ps-jblpm-ci}

Here we outline the MCMC algorithm for the DPMPM model, which was used for comparisons with the BLPM in Section 6 of the main text.

\begin{itemize}
 \item[\textit{Step 1:}] For $i=1,\dots,N$, sample $s_{i}\in \{1,\dots,K\}$ from a
   multinomial distribution with one trial and probabilities
\begin{align*}
\textrm{ Pr}(s_{i}=h|-)=\frac{\pi_{h}\rho_{h}^{W_i}(1-\rho_{h})^{1-W_i}\prod_{j=1}^{q}\psi_{h
    jZ_{ij}}}{\sum_{k=1}^{K}\pi_{k}\rho_{k}^{W_i}(1-\rho_{k})^{1-W_i}\prod_{j=1}^{q}\psi_{kjZ_{ij}}}.
\end{align*}

\item[\textit{Step 2:}] For $h=1,\dots, K-1$, sample $V_{h}$ from the Beta distribution
\begin{eqnarray*}
(V_{h}|-)\sim \textrm{Beta}(1+n_{h},\alpha+\sum_{k=h+1}^{K}n_{k}),
\end{eqnarray*}
where $n_{h}=\sum_{i=1}^{N}I(s_{i}=h)$.
Compute $\pi_{h}$ from
$\pi_{h}=V_{h}\prod_{k<h}(1-V_{k})$ with $V_K=1$.

\item[\textit{Step 3a:}] For $h=1, \dots, K$ and $j=1, \dots, q$, sample $\mathbf{\psi}_{hj}=(\psi_{hj1},\dots,\psi_{hjd_{j}})$ from
the Dirichlet distribution,
\begin{eqnarray*}
(\mathbf{\psi}_{hj}|-)\sim
\textrm{Dirichlet}(1+\sum_{i:s_{i}=h}I(Z_{ij}=1), \dots, 1+\sum_{i:s_{i}=h}I(Z_{ij}=d_{j})).
\end{eqnarray*}

\item[\textit{Step 3b:}] For $h=1, \dots, K$, sample $\rho_{h}$ from
  the Beta distribution,
\begin{eqnarray*}
(\rho_{h}|-)\sim \textrm{Beta}(1 + \sum_{i:s_{i}=h}I(W_{i}=1), 1 + \sum_{i:s_{i}=h}I(W_{i}=0)).
\end{eqnarray*}

\item[\textit{Step 4:}] Sample $\alpha$ from the Gamma distribution,
\begin{eqnarray*}
(\alpha|-)\sim \textrm{Gamma}(a_{\alpha}+K-1, b_{\alpha}-\mbox{log}\pi_{K}).
\end{eqnarray*}

\item[\textit{Step 5:}] For each $Z_{i,j}$ that is missing in the
  collected data,  sample a new value from the multinomial
  distribution,
\begin{eqnarray*}
&(Z_{i,j}|-) \sim \textrm{Multinomial}(\{1,\dots,d_{j}\},\psi_{s_{i}j1},\dots,\psi_{s_{i}jd_{j}}).
\end{eqnarray*}

\item[\textit{Step 6:}] Sample values of $W_{i}$ for units in the
  refreshment sample from the Bernoulli distribution,
\begin{eqnarray*}
(W_i|-) \sim \textrm{Bernoulli}(\rho_{s_i}).
\end{eqnarray*}

\end{itemize}

\section{Posterior Computation Algorithm for BLPM}
\label{ps-jblpm}

Here we outline the MCMC algorithm for the BLPM model described in
Section 4 of the main text.

\begin{itemize}
 \item[\textit{Step 1:}] For $i=1,\dots,N$, sample $s_{i}\in \{1,\dots,K\}$ from a
   multinomial distribution with one trial and probability $\textrm{Pr}(s_{i}\textrm{=}h|-)$ equal to
\begin{align*}
\frac{\pi_{h}\rho_{h}^{W_i}(1-\rho_{h})^{1-W_i} \prod_{j=1}^{q_0} \psi_{hjZ_{ij}}^{(W_i)}\prod_{j=q_0+1}^{q_0+q_1}\psi_{hjZ_{ij}} \prod_{j=q_0+q_1+1}^{q} \psi_{hjZ_{ij}}^{(W_i)}}
{\sum_{k=1}^{K}\pi_{k}\rho_{k}^{W_i}(1-\rho_{k})^{1-W_i} \prod_{j=1}^{q_0} \psi_{kjZ_{ij}}^{(W_i)}\prod_{j=q_0+1}^{q_0+q_1}\psi_{kjZ_{ij}} \prod_{j=q_0+q_1+1}^{q} \psi_{kjZ_{ij}}^{(W_i)}}.
\end{align*}

\item[\textit{Step 2:}] For $h\textrm{=}1,\dots, K-1$, sample $V_{h}$ from the Beta distribution,
\begin{eqnarray*}
(V_{h}|-)\sim \textrm{Beta}(1+n_{h},\alpha+\sum_{k\textrm{=}h+1}^{K}n_{k}),
\end{eqnarray*}
where $n_{h}\textrm{=}\sum_{i\textrm{=}1}^{N}I(s_{i}\textrm{=}h)$.
Compute $\pi_{h}$ from $\pi_{h}=V_{h}\prod_{k<h}(1-V_{k})$ with $V_K=1$.

\item[\textit{Step 3a:}] For $h\textrm{=}1, \dots, K$, $j\in \{1, \dots q_0, q_0+q_1+1, \dots, q\}$, sample $\mathbf{\psi}_{hj}^{(1)}\textrm{=}(\psi_{hj1}^{(1)},\dots,\psi_{hjd_{j}}^{(1)})$ from
the Dirichlet distribution,
\begin{eqnarray*}
(\mathbf{\psi}_{hj}^{(1)}|-)\sim \textrm{Dirichlet}(1+\sum_{i:s_{i}\textrm{=}h, w_i\textrm{=}1}I(Z_{ij}\textrm{=}1), \dots,  1 +\sum_{i:s_{i}\textrm{=}h, w_i\textrm{=}1}I(Z_{ij}\textrm{=}d_{j})).
\end{eqnarray*}

\item[\textit{Step 3b:}] For $h\textrm{=}1, \dots, K$, $j\in \{1, \dots q_0,  q_0+q_1+1, \dots, q\}$, sample  $\mathbf{\psi}_{hj}^{(0)}\textrm{=}(\psi_{hj1}^{(0)},\dots,\psi_{hjd_{j}}^{(0)})$ from
the Dirichlet distribution,
\begin{eqnarray*}
(\mathbf{\psi}_{hj}^{(0)}|-)\sim \textrm{Dirichlet}(1+\sum_{i:s_{i}\textrm{=}h, w_i\textrm{=}0}I(Z_{ij}\textrm{=}1),\dots, 1 +\sum_{i:s_{i}\textrm{=}h, w_i\textrm{=}0}I(Z_{ij}\textrm{=}d_{j})).
\end{eqnarray*}

\item[\textit{Step 3c:}] For $h=1, \dots, K$, $j\in \{q_0+1, \dots, q_0+q_1\}$, sample $\psi_{hj} = (\psi_{hj1},\dots,\psi_{hjd_{j}})$ from the
Dirichlet distribution,
\begin{eqnarray*}
(\mathbf{\psi}_{hj}|-)\sim \textrm{Dirichlet}(1+\sum_{i:s_{i}\textrm{=}h}I(Z_{ij}\textrm{=}1),\dots,1+\sum_{i:s_{i}\textrm{=}h}I(Z_{ij}\textrm{=}d_{j})).
\end{eqnarray*}

\item[\textit{Step 3d:}] For $h=1, \dots, K$, sample $\rho_{h}$ from the Beta distribution,
\begin{eqnarray*}
(\rho_{h}|-)\sim \textrm{Beta}(1+\sum_{i:s_{i}\textrm{=}h}I(W_{i}\textrm{=}1), 1+\sum_{i:s_{i}\textrm{=}h}I(W_{i}\textrm{=}0)).
\end{eqnarray*}
\item[\textit{Step 4:}] Sample $\alpha$ from the Gamma distribution,
\begin{eqnarray*}
(\alpha|-)\sim \textrm{Gamma}(a_{\alpha}+K-1, b_{\alpha}-\mbox{log}\pi_{K}).
\end{eqnarray*}

\item[\textit{Step 5a:}]  For  $j \in \{1, \dots q_0, q_0+q_1+1, \dots, q\}$, for each $Z_{ij}$ that is missing in the collected data in the panel or refreshment sample, sample a new value from the multinomial distribution,
\begin{eqnarray*}
(Z_{i,j}|-) \sim \textrm{Multinomial}(\{1,\dots,d_{j}\},\psi_{s_{i}j1}^{(w_i)},\dots,\psi_{s_{i}jd_{j}}^{(w_i)}).
\end{eqnarray*}

\item[\textit{Step 5b:}] For  $j \in \{q_0+1, \dots, q_0+q_1\}$, for each $Z_{ij}$ that is missing in the collected data in the panel or refreshment sample, sample a new value from the multinomial distribution,
\begin{eqnarray*}
(Z_{i,j}|-) \sim \textrm{Multinomial}(\{1,\dots,d_{j}\},\psi_{s_{i}j1},\dots,\psi_{s_{i}jd_{j}}).
\end{eqnarray*}
\item[\textit{Step 6:}] For all individuals in the refreshment sample, sample $W_{mis}$ from the Bernoulli distribution,
\begin{eqnarray*}
(W_i|-) \sim \textrm{Bernoulli}(\mbox{Pr}(W_{i}\textrm{=}1|Z_{i}, s_i, \Psi)),
\end{eqnarray*}
where $\mbox{Pr}(W_{i}\textrm{=}1|-)$ is equal to
\begin{align*}
\frac{f(X_{i}|s_i, W_i\textrm{=}1,\Psi)f(Y_{i,2}|s_i, W_i\textrm{=}1,\Psi)\mbox{Pr}(W_i\textrm{=}1|s_i, \Psi)}
{\sum_{w\in\{0,1\}}f(X_{i}|s_i, W_i\textrm{=}w,\Psi)f(Y_{i,2}|s_i, W_i\textrm{=}w,\Psi))\mbox{Pr}(W_i\textrm{=}w|s_i, \Psi)}.
\end{align*}

\end{itemize}


\section{Results of DPMPM on APYN Data}
\label{apyn-dpm}

The main text uses the BLPM for analysis of the APYN data. Here, we present some results using the DPMPM model. We set the initial values of parameters following similar strategies as those in Section 6.2.  We run the MCMC chain for 150,000 iterations, collecting $m\textrm{=}50$ imputed datasets after thinning. We implement some of the same analyses performed in Section 6.3. 

\begin{figure}[h!]
\centering
\begin{tabular}{c}	
	\includegraphics[width=\textwidth,right]{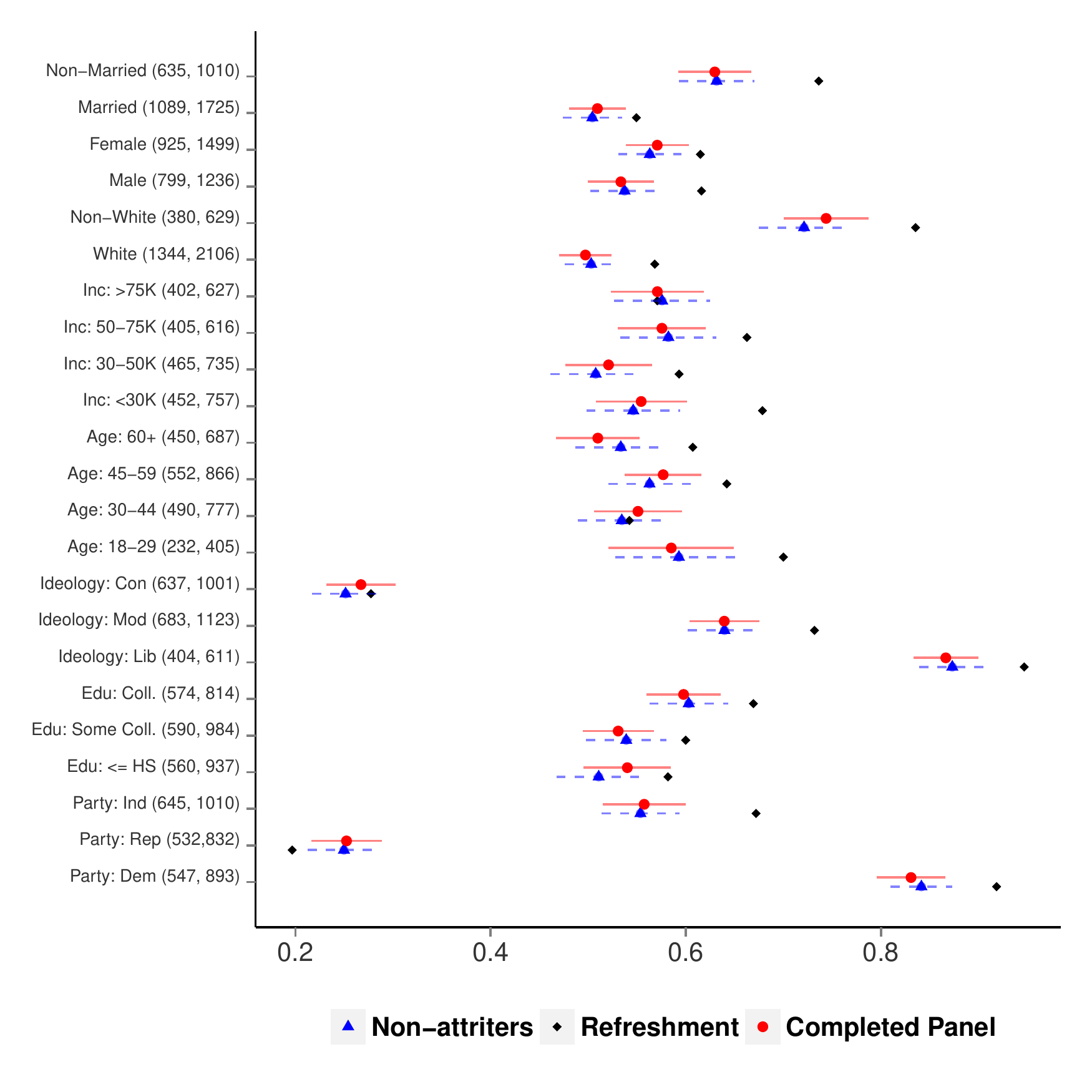}\\
\end{tabular}
\caption[Outputs of DPMPM on APYN data.]{Point estimates and 95\% confidence intervals for Obama favorability in various
   sub-groups. Results presented for the $N_{cp}$ panel non-attriters, the $N_r$ refreshment sample cases, and the $N_p$ panel participants. Inferences based on unweighted analyses of the $m\textrm{=}50$ completed datasets, after multiple imputation of missing values due to item nonresponse and attrition via the DPMPM model including $W$. The numbers in parentheses are the corresponding subgroup sizes, the first being the size among non-attriters and the second being among the completed panel. We randomly select one imputed dataset to obtain the sample sizes when the background variables are subject to item nonresponse.} 
\label{dpmpm-apyn}
\end{figure}

The DPMPM results in substantially different conclusions than those based on the BLPM, suggesting no bias due to attrition.  The marginal estimate for Obama favorability for panel non-attriters in the panel is 0.551, with 95\% confidence interval (0.527, 0.575).  The same estimate computed using only the panel attriters is 0.558 with 95\% confidence interval (0.509, 0.618).  Figure~\ref{dpmpm-apyn} displays the estimates of Obama favorability for the sub-groups in Figure 3 in the main article.  The completed panel estimates are similar to those for the non-attriters, so that the DPMPM does not appear to leverage the information in the refreshment sample. Apparently, the conditional independence assumption for $Y_2$ is inappropriate for
these data. 

We also estimated the DPMPM model without controlling for $W$, which effectively enforces a missing at random assumption. The results, not shown here for brevity, are very similar to those in Figure~\ref{dpmpm-apyn}.  

\begin{figure}
\begin{tabular}{c}
	\includegraphics[height=3.25in,width=0.85\textwidth,right]{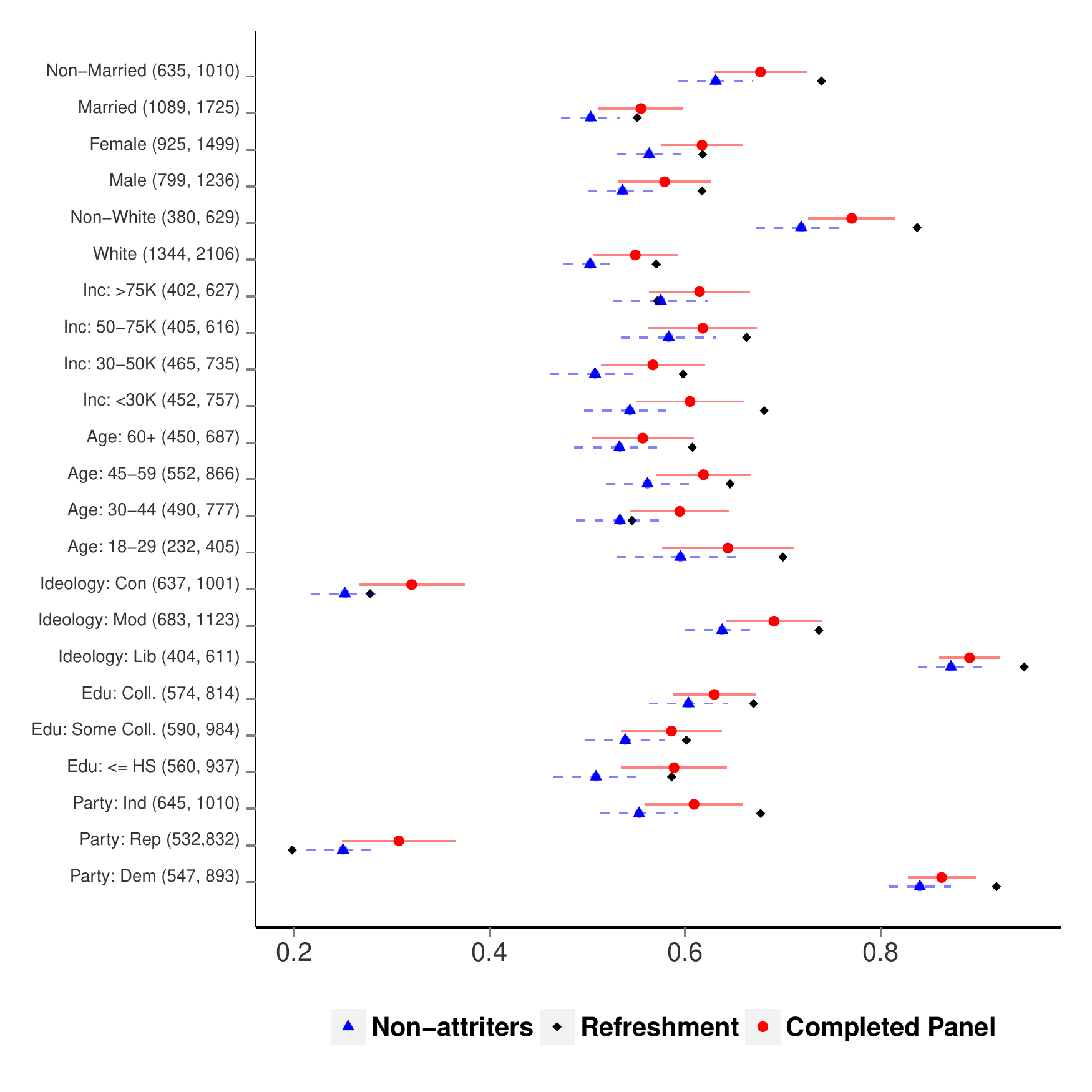}\\
	\includegraphics[height=3.25in,width=\textwidth,right]{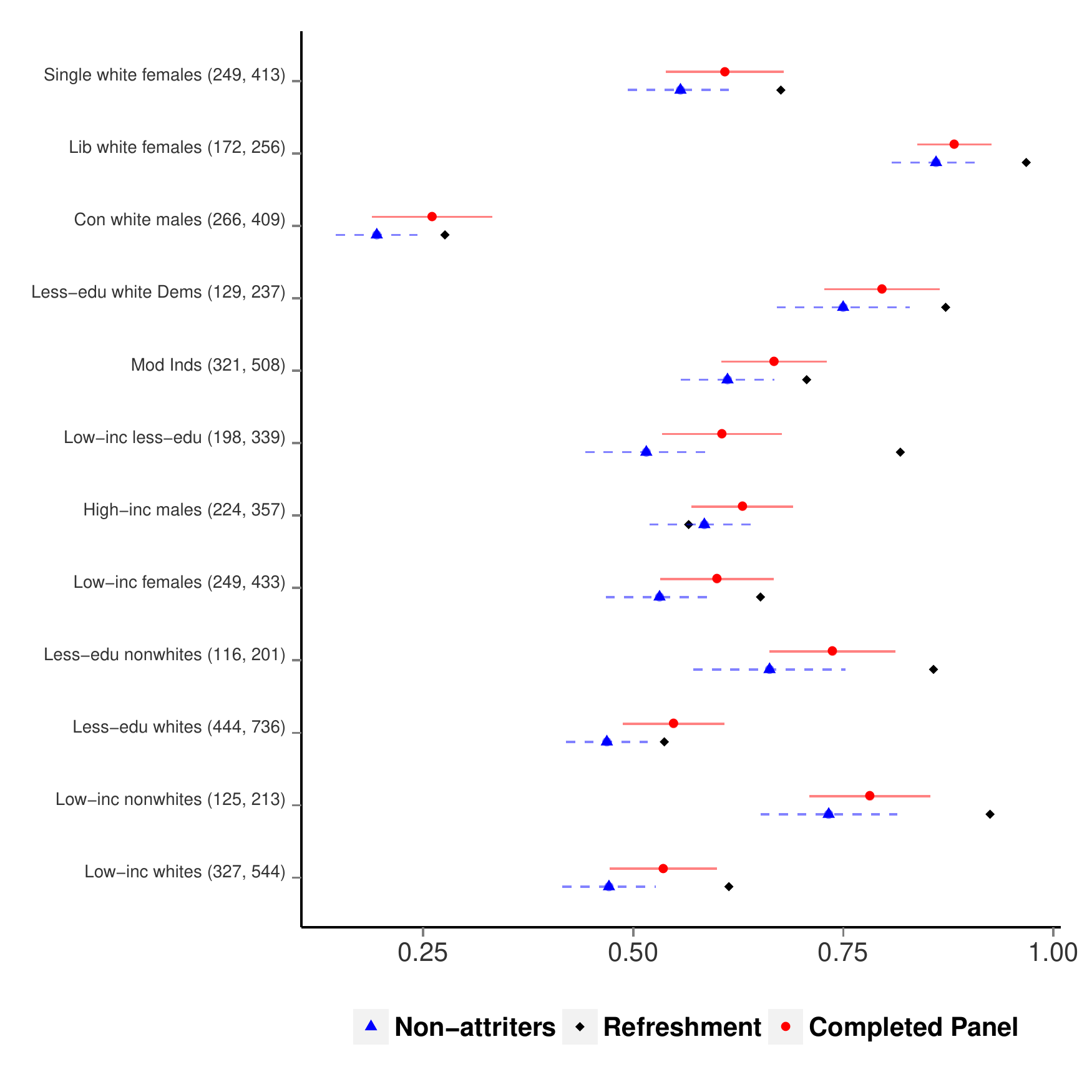}\\
\end{tabular}
 \caption[Conditional probability comparison on demographics between panel and refreshment samples]{Point estimates and 95\% confidence intervals for Obama favorability in sub-groups. Results presented for the $N_{cp}$ panel non-attriters, the $N_r$ refreshment samples, and the $N_p$ panel participants. Inferences based on unweighted analyses of the $m\textrm{=}50$ completed datasets, after multiple imputation of missing values via the AN model. The numbers in parentheses are the corresponding subgroup sizes, the first being the size among non-attriters and the second being among the completed panel. We randomly select one imputed dataset to obtain the sample sizes when the background variables are subject to item nonresponse. }
 \label{prob-x1-an}
\end{figure}

\section{Results of semi-parametric additive nonignorable model on APYN data}
\label{supp-ppp-an}

As described in Section 6.3 and 6.4 of the main text, we analyzed the APYN data using the semi-parametric additive 
non-ignorable (AN) model of \cite{si:reiter:hillygus:pa12}, including corresponding posterior predictive checks.
We set the initial values of parameters following similar strategies as those in Section 6.2
 and run the MCMC chain for 100,000 iterations. We create $m\textrm{=}50$ completed datasets that we use for 
the analyses described in Section 6.3. 

Like the BLPM, the semi-parametric AN model also indicates that attrition is non-ignorable. The
 estimated coefficient for $Y_2$ in the selection model for $W$ is significantly different from zero.  Comparing the substantive 
analyses done for the BLPM in  Figure 3 and 4 of the main text, we find that 
the  semi-parametric AN model yields similar results. 
Figure~\ref{prob-x1-an} displays estimated probabilities for Obama favorability for each of the sub-groups 
defined by the time-invariant variables.  Like the BLPM, the semi-parametric AN model appears to correct for attrition bias. 
The point estimates for BLPM and the semi-parametric AN model differ a bit; however, the large variances after multiple imputation 
 make the comparison practically indistinguishable.

\begin{figure}[h!]
\centering
\begin{tabular}{c}
	\includegraphics[width=\textwidth,height=2in]{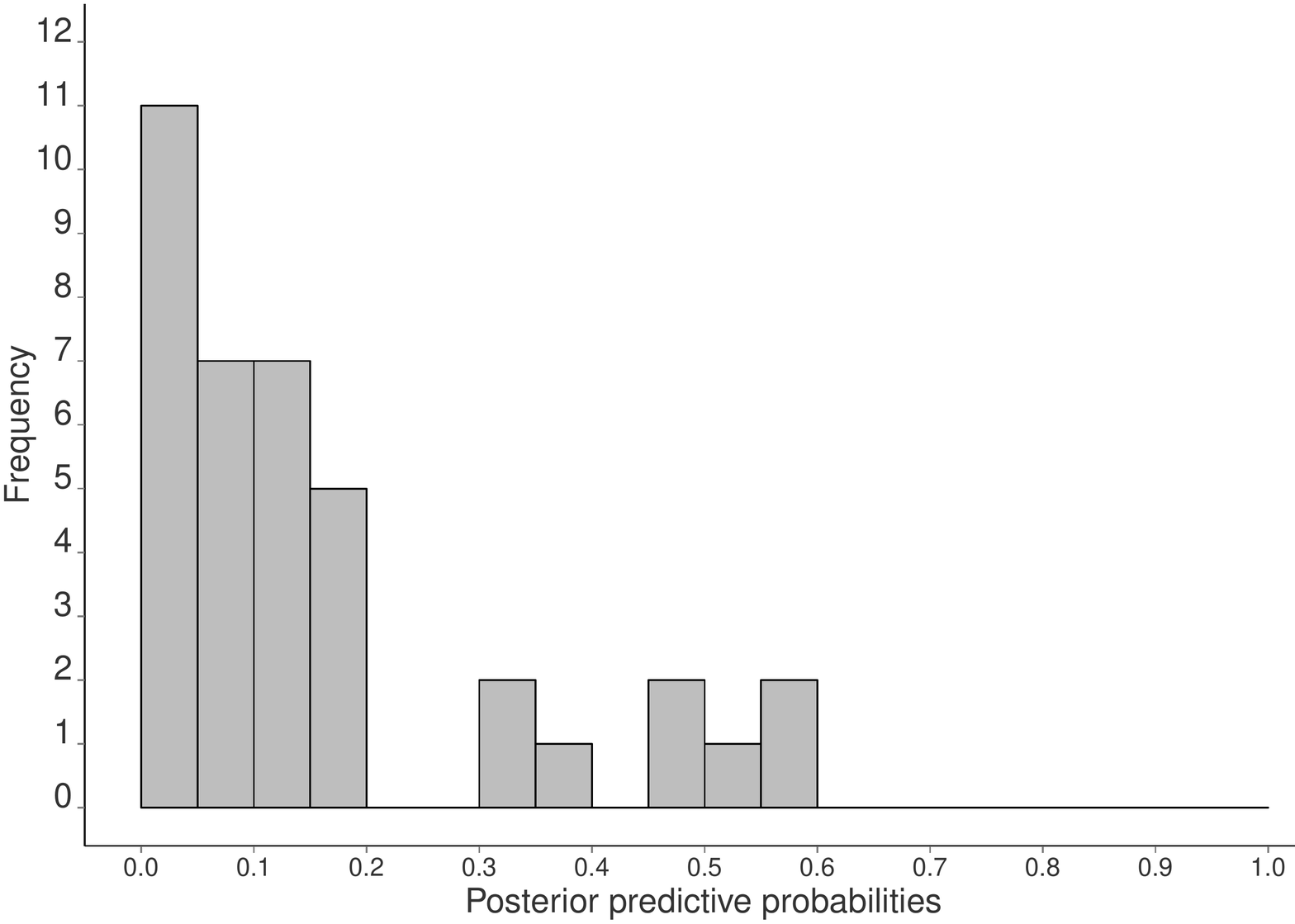}\\
\end{tabular}
\caption[Two-sided ppp for AN]{Histogram of the 38 $ppp$ values from the posterior predictive checks of the APYN analyses with AN model. }
\label{ppp2}
\end{figure}

We also examined posterior predictive probabilities, as noted in Section 6.4 of the main text.
Figure~\ref{ppp2} displays a histogram of the 38 values of $ppp$ based on the semi-parametric  
AN model. Many $ppp$ values are quite small, suggesting that overall the semi-parametric AN model does 
not to fit the APYN data particularly well, especially compared to the BLPM.  

\section{Frequency distribution for the posterior predictive probabilities for BLPM model}
\label{supp-ppp}
\begin{figure}[h!]
\centering
\begin{tabular}{c}
	\includegraphics[height=\textwidth,width=2in, angle=-90]{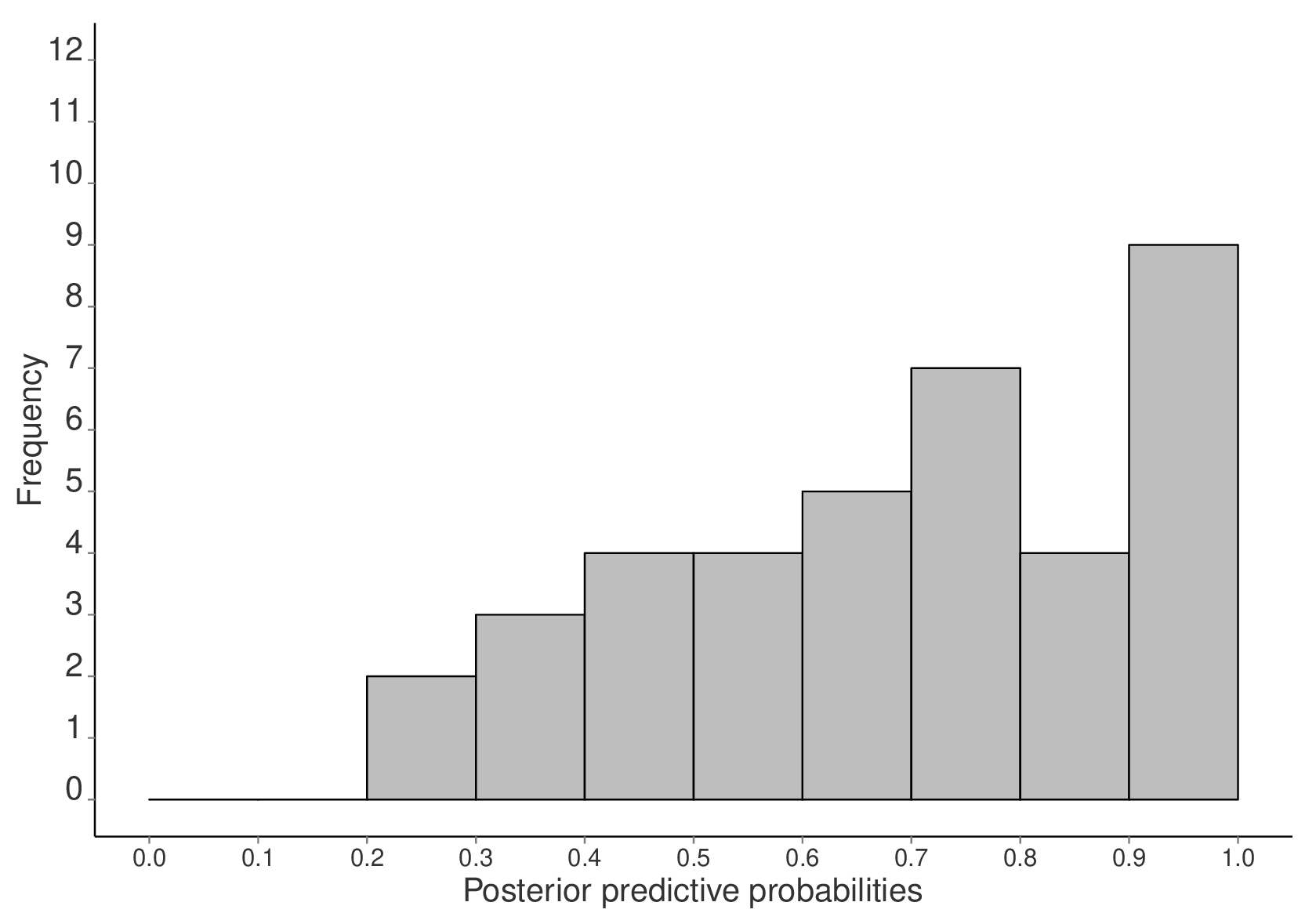}\\
\end{tabular}
\caption[Two-sided ppp for BLPM]{Histogram of the 38 $ppp$ values from the posterior predictive checks of the APYN analyses with BLPM model. }
\label{ppp}
\end{figure}

Figure~\ref{ppp} displays  a histogram of the 38 values of $ppp$ with BLPM, as described in Section 6.4 of the main text.
 None of $ppp$ values are below $0.20$, suggesting no evidence of serious lack of model fit.

\bibliographystyle{chicago}
\bibliography{blpm}